\documentclass[twocolumn,noshowpacs,nofootinbib,amsmath,amssymb]{revtex4}

\usepackage{graphicx}
\usepackage{amssymb,amsmath}
\usepackage{bm}

\begin{document}

\title{Isospin symmetry breaking in double-pion production \\ in the region of $\bm{d^*(2380)}$ and the scalar $\bm{\sigma}$ meson}
\author{\firstname{M.~N.}~\surname{Platonova}}
\email{platonova@nucl-th.sinp.msu.ru}
\author{\firstname{V.~I.}~\surname{Kukulin}}
\thanks{Deceased}
\affiliation{Skobeltsyn Institute of Nuclear Physics, \\
Lomonosov Moscow State University, Moscow 119991, Russia}
%
%


\begin{abstract}
The first attempt is made to provide a quantitative theoretical
interpretation of the WASA-at-COSY experimental data on the basic
double-pion production reactions $pn \to d \pi^0\pi^0$ and $pn \to
d \pi^+\pi^-$ in the energy region $T_p =1$--$1.3$~GeV [P.
Adlarson et al., Phys. Lett. B 721, 229 (2013)]. The data are
analyzed within a model based on production and decay of an
intermediate $I(J^P)=0(3^+)$ dibaryon resonance
$\mathcal{D}_{03}$ (denoted also as $d^*(2380)$). The observed decrease of the
near-threshold enhancement (the so-called ABC effect) in the
reaction $pn \to d \pi^+\pi^-$ in comparison to that in the reaction $pn
\to d \pi^0\pi^0$ is explained (at least partially) to be due to isospin symmetry
violation in the two-pion decay of an intermediate near-threshold
scalar $\sigma$ meson emitted from the $\mathcal{D}_{03}$ dibaryon
resonance under conditions of the partial chiral symmetry restoration.
\end{abstract}


\maketitle

\newpage

\section{Introduction}
\label{sec-intro}

The puzzling $\sigma$ meson (denoted as $f_0(500)$ in the latest Review of Particle Physics~\cite{PDG20}) has been the subject of an active discussion in hadronic physics for almost 60 years. The reason for this active interest is not only the complicated and still unclear nature of
the $\sigma$ meson or its very large width, which complicates
considerably determination of its basic properties, but mainly
the fundamental role of this lightest scalar meson in nuclear
physics and in physics of the strong interaction at all (see the dedicated review~\cite{Pelaez16}). There is extensive literature about the role of the $\sigma$ meson in the chiral symmetry breaking and restoration in QCD. Being the
lowest resonance in QCD with the vacuum quantum numbers, the
$\sigma$ meson may be responsible for the constituent quark masses,
hence, it is sometimes called the Higgs particle of the strong
interaction~\cite{Schum10}. Furthermore, the $\sigma$ meson
is traditionally considered as a basic agent of the
intermediate-range $NN$ attraction in nuclei, which is very
difficult to reconcile with its huge width and, consequently, a
very short lifetime~\cite{YAF2013}. Besides that, the dynamics of $\sigma$-meson production in hadronic processes is poorly understood to date. Production of $\sigma$ mesons from Pomerons has been predicted theoretically in high-energy $pp$ collisions~\cite{Kisslinger05}. Experimentally, huge accumulation of data at the $\pi\pi$ invariant masses of 400--500 MeV has been observed in double-pion
production in $pp$ collisions at energies $E \gtrsim 100$~GeV~\cite{Alde97} and also in quarkonia decays~\cite{Ablikim07}, which can likely be explained by production of light scalar mesons. Other indications of $\sigma$-meson production in hadronic processes can be found, e.g., in the review~\cite{Pelaez16}. For the above reasons, studying the
possibilities to observe $\sigma$-meson production in hadronic
collisions at high and intermediate energies is of great interest. In this paper, we analyze the
indications of the intermediate $\sigma$-meson generation in
double-pion production in $pn$ collisions in the GeV
energy region.

In a series of experiments performed by the WASA-at-COSY
Collaboration in Juelich FZ, the first high-statistics exclusive
data on a number of double-pion production reactions in $pn$, $pd$,
and $dd$ collisions at energies $T_p =0.8$--$1.4$~GeV have been
obtained (see Refs.~\cite{Adl11,Adl13-iso,Adl15-pd,Adl12-dd} and reviews~\cite{Clem17,Clem21}). In these experiments, existence
of the near-threshold enhancement in the $\pi\pi$ invariant-mass
spectrum known since 1960s as the Abashian--Booth--Crowe (ABC) effect~\cite{ABC} in
all reactions accompanied by isoscalar dipion production and
formation of the bound nuclei in the final state has been
confirmed. The same experiments revealed for the first time
generation of a dibaryon resonance $\mathcal{D}_{03}$ (or $d^*$)
with quantum numbers $I(J^P)=0(3^+)$, the mass
$M_{D_{03}} \simeq 2.38$ GeV and the rather narrow width
$\Gamma_{D_{03}} \simeq 70$ MeV and a direct correlation
of this resonance with the ABC effect. After that, this resonance
has been confirmed by the partial-wave analysis (PWA) of $np$ elastic scattering, which included the new data in the region of the $\mathcal{D}_{03}$ excitation~\cite{Adl14-el}. So, among a large
number of dibaryon resonances proposed for the last 50 years since
the first theoretical prediction of dibaryons by Dyson and
Xuong~\cite{Dyson64}, the $\mathcal{D}_{03}$ resonance is the most
reliably established to date. However the detailed mechanism of
the $\mathcal{D}_{03}$ decay which leads to the ABC enhancement
is still a subject of debates~\cite{Bash17-ABC}. In particular, the decay $\mathcal{D}_{03} \to
\Delta\Delta$ proposed in~\cite{Adl11} gives a strong
near-threshold enhancement only under an assumption of a
very soft vertex form factor, which is hardly compatible with the known properties (mass,
width, and size) of the $\mathcal{D}_{03}$ resonance.

In Ref.~\cite{PRC2013} we proposed a model for the
basic double-pion production reaction $pn \to d\pi^0\pi^0$, which
included two mechanisms: $pn \to \mathcal{D}_{03} \to d\sigma \to
d \pi^0\pi^0$ and $pn \to \mathcal{D}_{03} \to
\mathcal{D}_{12}\pi^0 \to d \pi^0\pi^0$, where $\mathcal{D}_{12}$
is the known isovector dibaryon resonance with the quantum numbers $I(J^P)=1(2^+)$, the mass
$M_{D_{12}} \simeq 2.15$ GeV and the width
$\Gamma_{D_{12}} \simeq 110$ MeV (see, e.g., Ref.~\cite{Hoshizaki93}). By taking into
account two above interfering decay routes for the
$\mathcal{D}_{03}$ resonance, we explained very well the total
cross section, the $\pi\pi$ and $d\pi$ invariant-mass spectra and
the final pion and deuteron angular distributions in the $pn \to
\pi^0\pi^0$ reaction near the resonance peak energy $T_p=1.14$ GeV
(or $\sqrt{s}=2.38$ GeV)~\cite{PRC2013,NPA2016,FBS2014}. The ABC enhancement in our model has been interpreted as a consequence of a scalar $\sigma$-meson emission which has lower mass and width than the respective free-space values~\cite{PDG20} due to the partial chiral symmetry restoration in the excited $\mathcal{D}_{03}$ dibaryon. Such an interpretation finds support in a number of theoretical and experimental works (see, e.g., the studies of the chiral symmetry restoration in excited hadrons~\cite{Glozman00,Glozman07}, the observation of the very light $\sigma$ mesons in $d{\rm C}$ collisions~\cite{Abraamyan}, and other related references in the review~\cite{AP10K}), though more confirmations of light scalar meson production just in the dibaryon decays are still needed.

The model~\cite{PRC2013} is straightforwardly applicable to the reaction $pn \to
d \pi^+\pi^-$, for which the exclusive data on total and
differential distributions have been obtained in a recent
experiment~\cite{Adl13-iso}. These measurements have revealed the
significant isospin symmetry violation in the $\pi\pi$ invariant-mass spectra in the near-threshold region,
i.e., a suppression of the ABC enhancement by about 25\% for
charged dipion production as compared to neutral dipion
production. The authors~\cite{Adl13-iso} supposed this suppression
to be due to a phase-space reduction originating from the 5-MeV
mass difference between the charged and neutral pions. However, no
quantitative theoretical interpretation of these data has been
published to date. Since the observed effect is large and energy-dependent,
some additional sources for isospin symmetry breaking
related to the near-threshold reaction dynamics should be examined
as well. It is worth mentioning that a similar isospin symmetry violation effect was obtained in an earlier CELSIUS-WASA experiment~\cite{Bash06} on the reactions $pd \to {}^3{\rm He}\pi\pi$ as well as $pp \to pp\pi\pi$. There, it was claimed to be due to the pion loops which enhance the $\pi^0\pi^0$ production cross section below the $\pi^+\pi^-$ threshold~\cite{Bash06}. However, no explicit calculations were performed to confirm this claim.

The purpose of the present paper is to analyze the impact of the different reaction mechanisms involving the $\mathcal{D}_{03}$ dibaryon production and decay on the observed isospin symmetry violation
in the reactions $pn \to d\pi^0\pi^0$ and
$pn \to d\pi^+\pi^-$ near the two-pion threshold. We will try to give a description of the
experimental data~\cite{Adl13-iso} on both reactions within an updated model, which combines two mechanisms proposed previously in Ref.~\cite{PRC2013} and a mechanism of the $\mathcal{D}_{03}$ decay into two intermediate $\Delta$-isobars suggested in Ref.~\cite{Adl11}. Adding the $\mathcal{D}_{03} \to \Delta\Delta$ mechanism is needed to take into account the restriction on the $\mathcal{D}_{12}\pi$ decay mode of the $\mathcal{D}_{03}$ dibaryon, imposed by the recent experimental data on the isoscalar $NN \to NN\pi$ cross section~\cite{Adl17-NNpi}. Another necessary modification of the model~\cite{PRC2013} concerns the accurate treatment of the pion mass difference in the $\sigma$-meson decay, which is important just for the description of near-threshold double-pion production in both $\pi^0\pi^0$ and $\pi^+\pi^-$ channels. It will be shown that the observed difference in the ABC enhancements in the above two reactions can be
explained (at least partially) by the isospin symmetry breaking in the decay of the intermediate scalar $\sigma$ meson, which is
shifted towards the two-pion threshold by the partial chiral symmetry
restoration in the excited $\mathcal{D}_{03}$ dibaryon.

The paper is organized as follows. In Sec.~\ref{sec-form}, we outline the basic formalism of the model used in the calculations of the reactions $pn \to d (\pi\pi)_0$ (where the subscript ``0'' means the isoscalar $\pi^0\pi^0$ or $\pi^+\pi^-$ state) in the region of $T_p = 1$--$1.3$ GeV corresponding to the $\mathcal{D}_{03}$ resonance excitation. In Sec.~\ref{sec-dd}, we study the two-pion invariant-mass spectra at $\sqrt{s}=2.38$ GeV, which result from the $\mathcal{D}_{03}$ decay via the intermediate $\mathcal{D}_{12}\pi$ or $\Delta\Delta$ states, with regard to the impact of each of these mechanisms on the observed isospin symmetry breaking in $pn$-induced $\pi\pi$ production. In the next Sec.~\ref{sec-sig}, we study the contribution of the $\mathcal{D}_{03} \to d\sigma$ decay mode and combine it with both above mechanisms to analyze the predictions of the full model. Sec.~\ref{sec-ed} is dedicated to the analysis of the energy dependence of the observed isospin symmetry violation in double pion production. We summarize our results in Sec.~\ref{sec-sum}.

\section{Formalism for the double-pion production reactions $\bm{pn \to d (\pi\pi)_0}$ in the region of $\bm{d^*(2380)}$}
\label{sec-form}

As in Refs.~\cite{PRC2013,NPA2016}, we consider the $\mathcal{D}_{03}$ (or $d^*(2380)$) dibaryon as a hexaquark-dominated state with a basic structure $4q$--$2q$, where the tetraquark $4q$ ($ST=10$) and diquark $2q$ ($S'T'=00$) clusters are connected by a color string with an orbital excitation $L=2$ (and a small admixture of $L=4$). So, we agree qualitatively with the microscopic quark-model calculations~\cite{Huang15,Huang16}, where the mass and narrow width of the $\mathcal{D}_{03}$ are explained by the dominance of the ``hidden color'' six-quark component suggested first in Ref.~\cite{BBC13}.\footnote{Recently, the three-diquark models for the $\mathcal{D}_{03}$ state have been also proposed~\cite{Shi19,Oka20}.} As we proposed in Refs.~\cite{PRC2013,NPA2016}, such a state can decay {\emph directly} into another six-quark state by meson emission, in a full analogy with ordinary excited hadrons. Thus, the $\mathcal{D}_{03}$ dibaryon can emit the scalar $\sigma$-meson (in $d$ wave) and de-excite into the $\mathcal{D}_{01}$ state, which has the same structure as the $\mathcal{D}_{03}$ but with the $L=0$ string between the $4q$ and $2q$ clusters (with a small admixture of $L=2$). The $\mathcal{D}_{01}$ is nothing else than the six-quark (or dibaryon) component of the deuteron, which becomes a physical deuteron in the course of dressing by the $NN$ loops. Alternatively, the $\mathcal{D}_{03}$ dibaryon can emit sequentially two $p$-wave pions and also come to the final deuteron via an intermediate isovector dibaryon state $\mathcal{D}_{12}$ strongly coupled to the $S$-wave $N\Delta$ state. So, the $\mathcal{D}_{03}$ resonance is an analog of the Roper resonance $N^*(1440)$ by its two-pion decay modes. 

However, the $\mathcal{D}_{03}$ state can also contain $3q$--$3q$ configurations with colorless three-quark clusters, the main of which is the $\Delta\Delta$ configuration.
In fact, all known six-quark states are located near the di-hadron thresholds and should be strongly coupled to the respective di-hadron molecular-like states~\cite{Clem21,EPJA20S,Clem20}. The nearer the threshold, the stronger the coupling, and the higher the weight of the respective molecular-like state in the total dibaryon wavefunction. Contrary to the deuteron and the $\mathcal{D}_{12}$ dibaryon, the $\mathcal{D}_{03}$ state is located rather far from the respective (in this case, the $\Delta\Delta$) threshold, so the molecular-like $\Delta\Delta$ state has a small fraction in it. Hence, when we talk about the $\Delta\Delta$ component of the $\mathcal{D}_{03}$ dibaryon, we should mean predominantly the compact six-quark configuration consisting of two three-quark clusters with the quantum numbers of the $\Delta$, which are in the relative $S$ wave, so that, the two $\Delta$ baryons are overlapped strongly in such a state. According to the calcuations~\cite{Huang15,Huang16}, the $\Delta\Delta$ component constitutes about 30\% of the total $\mathcal{D}_{03}$ wavefunction.\footnote{This fraction should be further examined, however, due to cautions need to be taken in explaining the configuration structure for any baryon-baryon bound system in the case that the wavefunction of a single baryon is
not consistent with the given Hamiltonian~\cite{Huang18}.} Just this component can decay directly into two physical $\Delta$-isobars. In Refs.~\cite{PRC2013,NPA2016}, we neglected this decay mode of the $\mathcal{D}_{03}$ dibaryon, while it is considered to be dominant in many other works (see the review~\cite{Clem17}). We considered the $\mathcal{D}_{03}$ decay into the $\mathcal{D}_{12}+\pi$ intermediate state to be the dominant one, based also on the fact that this state has the two times smaller width and, accordingly, two times longer lifetime than the $\Delta\Delta$ intermediate state. However, the recent experiment~\cite{Adl17-NNpi} revealed a very small, if any, $NN\pi$ decay branch of the $\mathcal{D}_{03}$ resonance, thus imposing restrictions on its decay via the $\mathcal{D}_{12}+\pi$ intermediate state. So, for consistency with the data~\cite{Adl17-NNpi}, we should reduce the contribution of this decay mode and include some admixture of the $\Delta\Delta$ one (see details at the end of this section).
It should be also noted that the observed branching ratios (BR) for the $\mathcal{D}_{03}$ decay into different final states~\cite{Bash15-BR} can be reproduced by some combination of the $\mathcal{D}_{12}\pi$ and $\Delta\Delta$ components in the $\mathcal{D}_{03}$ state~\cite{Gal17}. The data~\cite{Adl17-NNpi} restrict the $\mathcal{D}_{12}\pi$ component to be not more than 25\%. However, we do not consider the $\mathcal{D}_{12}+\pi$ state as a component of the $\mathcal{D}_{03}$ resonance, but rather as an intermediate state in the decay of its dominating six-quark component.

\begin{figure}[!ht]
\begin{center}
\resizebox{1.0\columnwidth}{!}{\includegraphics{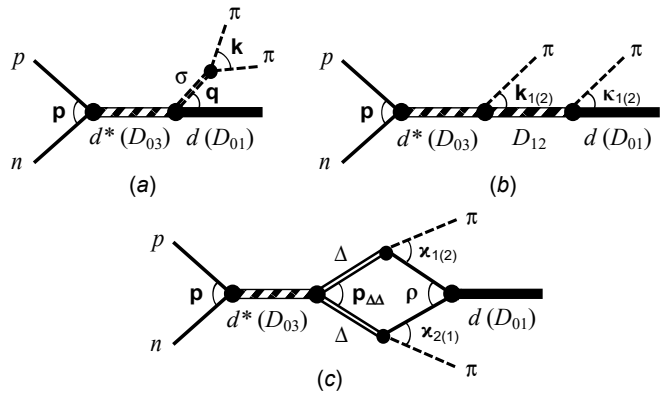}}
\end{center}
\caption{Diagrams of different mechanisms for double-pion production in the region of the $\mathcal{D}_{03}$ (or $d^*(2380)$) resonance formation. The $3$-momenta in the pair center-of-mass frames are indicated between the respective lines.}
\label{fig-dia}
\end{figure}

Thus, we take into account the
$\mathcal{D}_{03}(2380)$ dibaryon formation in a $pn$
collision and its subsequent decay via three interfering routes: (i)
through emission of the intermediate scalar $\sigma$-meson, which decays into
two final pions in the scalar-isoscalar channel, (ii) through the intermediate isovector $\mathcal{D}_{12}(2150)$ dibaryon
production, which decays in turn into a pion and a final
deuteron, and (iii) through the intermediate $\Delta\Delta$ state, which decays into two pions and two nucleons merging finally into a
deuteron. These mechanisms for double-pion production are depicted in Fig.~\ref{fig-dia}. We do not include here the $t$-channel background processes of $NN^*(1440)$ or $\Delta\Delta$ excitation, since our study is focused on double-pion production in the vicinity of the $\mathcal{D}_{03}$ peak, where these processes should give a small contribution~\cite{Adl11}. Their inclusion would also enlarge the number of adjustable parameters.

The amplitude for the reactions $pn \to d(\pi\pi)_0$ with
account of the above three mechanisms can be written as follows:
\begin{equation}
\label{amp} \mathcal{M}_{\lambda_p,\lambda_n,\lambda_d} =
\frac{\sum\limits_{\lambda_3}
\mathcal{M}^{(D_{03})}_{\lambda_p,\lambda_n,\lambda_3}\left[\mathcal{M}^{(\sigma)}_{\lambda_3,\lambda_d}
+
\mathcal{M}^{(D_{12})}_{\lambda_3,\lambda_d}
+ \mathcal{M}^{(\Delta\Delta)}_{\lambda_3,\lambda_d}\right]}{s-M_{D_{03}}^2+i\sqrt{s}\Gamma_{D_{03}}(s)},
\end{equation}
where $\mathcal{M}^{(D_{03})}_{\lambda_p,\lambda_n,\lambda_3}$ stands for the helicity amplitude of the $\mathcal{D}_{03}$ dibaryon formation and $\mathcal{M}^{(\sigma)}_{\lambda_3,\lambda_d}$, $\mathcal{M}^{(D_{12})}_{\lambda_3,\lambda_d}$, and $\mathcal{M}^{(\Delta\Delta)}_{\lambda_3,\lambda_d}$ stand for the helicity amplitudes of its decay via three above routes.

When choosing the $z$ axis to be parallel to the initial center-of-mass
momentum ${\bf p}$, the $\mathcal{D}_{03}$ dibaryon formation
amplitude takes the form~\cite{NPA2016}
\begin{equation}
\mathcal{M}^{(D_{03})}_{\lambda_p,\lambda_n,\lambda_3} =
F_{pn \to D_{03}}(p)
C^{3\lambda_3}_{1\lambda_320}C^{1\lambda_3}_{\frac{1}{2}\lambda_p\frac{1}{2}\lambda_n}Y_{20}({\hat p}),
\label{ampd03}
\end{equation}
where $C^{J\Lambda}_{s_1\lambda_1 s_2\lambda_2}$ are the
Clebsch--Gordan coefficients. In turn, for the dibaryon decay
amplitudes, one gets the following expressions~\cite{NPA2016}:
\begin{equation}
\mathcal{M}^{(\sigma)}_{\lambda_3,\lambda_d} = \frac{F_{D_{03} \to
d \sigma}(q) F_{\sigma \to
\pi\pi}(k)}{M_{\pi\pi}^2-m_{\sigma}^2+iM_{\pi\pi}\Gamma_{\sigma}(M_{\pi\pi}^2)}
C^{3\lambda_3}_{1\lambda_d 2\mu}Y_{2\mu}({\hat q}),
\label{ampsig}
\end{equation}
\begin{equation*}
\mathcal{M}^{(D_{12})}_{\lambda_3,\lambda_d} =
\frac{F_{D_{03} \to D_{12} \pi_1}(k_1) F_{D_{12} \to d
\pi_2}(\kappa_1)}{M_{d\pi_2}^2-M_{D_{12}}^2+iM_{d\pi_2}\Gamma_{D_{12}}(M_{d\pi_2}^2)}
\end{equation*}
\begin{equation}
\times \sum\limits_{\lambda_2} \!\! C^{3\lambda_3}_{2\lambda_2 1\mu_2} C^{2\lambda_2}_{1\lambda_d 1\mu_1}Y_{1\mu_2}({\hat k_1})Y_{1\mu_1}({\hat \kappa_1}) + (\pi_1 \leftrightarrow \pi_2),
\label{ampd12}
\end{equation}
where we introduced the center-of-mass frame momenta of the final deuteron $\bf q$ and pions $\bf k_i$ ($i=1,2$), as well as the pion momenta ${\bm \kappa}_i$ ($i=1,2$) in the center-of-mass frame of the $i$-th pion and the deuteron. From the properties of the Clebsch--Gordan coefficients, one gets for the projections of the orbital angular momenta, which appear in Eqs.~(\ref{ampsig}) and (\ref{ampd12}): $\mu = \lambda_3 - \lambda_d$, $\mu_1 = \lambda_2 - \lambda_d$ and $\mu_2 = \lambda_3 - \lambda_2$. After taking explicitly the sum over $\lambda_2$ in Eq.~(\ref{ampd12}), the angular part of the amplitude (\ref{ampd12}) takes a form very similar to that of the amplitude (\ref{ampsig}) (see details in Ref.~\cite{NPA2016}).

In the present work we consider also the $\mathcal{D}_{03}$ decay into two $\Delta$-isobars (the so-called $s$-channel $\Delta\Delta$ mechanism). The formulas for the respective amplitude have been given in Ref.~\cite{Bash17-ABC}. We use here essentially the same formulas to calculate the contribution of the $s$-channel $\Delta\Delta$ mechanism but we also take into account the nucleon recoil in the $\Delta \to N \pi$ vertices,
which was neglected in~\cite{Bash17-ABC}. So, we take the amplitude for the process $\mathcal{D}_{03} \to \Delta\Delta \to d \pi\pi$ in the form
\begin{equation*}
\mathcal{M}^{(\Delta\Delta)}_{\lambda_3,\lambda_d} = \int \frac{d^3\rho}{(2\pi)^3} \, \varphi_d(\rho) F_{D_{03} \to
\Delta\Delta}(p_{\Delta\Delta})
\end{equation*}
\begin{equation*}
\times G_{\Delta}(M_{N_1\pi_1}) G_{\Delta}(M_{N_2\pi_2}) F_{\Delta \to N_1
\pi_1}(\varkappa_1) F_{\Delta \to N_2 \pi_2}(\varkappa_2)
\end{equation*}
\begin{equation*}
\times \sum\limits_{\lambda_{\Delta_1}\lambda_{N_1}} C^{3\lambda_3}_{\frac{3}{2}\lambda_{\Delta_1}\frac{3}{2}\lambda_{\Delta_2}}C^{\frac{3}{2}\lambda_{\Delta_1}}_{\frac{1}{2}\lambda_{N_1}1\mu_1}
C^{\frac{3}{2}\lambda_{\Delta_2}}_{\frac{1}{2}\lambda_{N_2}1\mu_2} C^{1\lambda_d}_{\frac{1}{2}\lambda_{N_1}\frac{1}{2}\lambda_{N_2}}
\end{equation*}
\begin{equation}
\times
Y_{1\mu_1}(\hat \varkappa_1)Y_{1\mu_2}(\hat \varkappa_2) + (\pi_1 \leftrightarrow \pi_2),
\label{ampdd}
\end{equation}
where
$\bf \varkappa_i$ is the pion momentum in the $\Delta_i$ rest frame ($i=1,2$), $G_{\Delta}(M_{N_i\pi_i}) = [M_{N_i\pi_i}^2-m_{\Delta}^2+iM_{N_i\pi_i}\Gamma_{\Delta}(M_{N_i\pi_i}^2)]^{-1}$ is the $\Delta$ propagator and $\varphi_d(\rho)$ is the deuteron wavefunction. We neglected the $D$-wave state of the deuteron, as in Ref.~\cite{Bash17-ABC}. In the calculations presented below, we used the $S$-wave component of the CD-Bonn wavefunction~\cite{Machl01}.

From the total amplitude $\mathcal{M}_{\lambda_p,\lambda_n,\lambda_d}$ defined by Eq.~(\ref{amp}),
one can find the $\pi\pi$ invariant-mass distribution:
\begin{equation}
\label{dsdm} \frac{d\sigma}{dM_{\pi\pi}} =  \frac{1}{(4\pi)^{5} p s}\!
\int{\!\! \int{\! qk d\Omega_{\hat q} d\Omega_{\hat k} \,
\overline{|\mathcal{M}_{\lambda_p,\lambda_n,\lambda_d}({\bf q},{\bf k})|^2}}},
\end{equation}
where $s$ is the total invariant energy, ${\bf k}$ is the pion momentum in center-of-mass frame of two pions, and the line over the matrix element squared stands for averaging over the initial and summing over the final spin states.

Then one gets for the total cross section:
\begin{equation}
\label{st} \sigma = \int\limits_{2m_{\pi}}^{\sqrt{s}-m_d} \!\! {d
M_{\pi\pi} \frac{d\sigma}{dM_{\pi\pi}}}.
\end{equation}

The vertex functions introduced in Eqs.~(\ref{ampd03})--(\ref{ampdd}) are related to
the partial decay widths as\footnote{The factor $\sqrt{4\pi}$ in the following relation has been canceled by the factor $1/\sqrt{4\pi}$ in the definition of the spherical harmonics $Y_{lm}$ entering the angular parts of the respective amplitudes. So, the $Y_{lm}$ without this factor are used in Eqs.~(\ref{ampd03})--(\ref{ampdd}). Note also that the factor $p^l$ is here included in the vertex function $F_{R \to ab}(p)$, while in Ref.~\cite{NPA2016} it was included in the (solid) spherical harmonics.}
\begin{equation}
F_{R \to ab}(p) = M_{ab}\sqrt{\frac{8\pi\Gamma_{R \to
ab}(p)}{p}},
\end{equation}
where $p$ is the momentum of the particle $a$ in the center-of-mass frame of the particles $a$ and $b$, related to the invariant mass as usual: $p = [(M_{ab}^2-m_a^2-m_b^2)^2-4m_a^2m_b^2]^{1/2}\big/2M_{ab}$, and $l$ is the relative orbital angular momentum of the particles $a$ and $b$.

For the partial decay widths with meson emission, we use the standard parametrization 
\begin{equation}
\label{G} \Gamma_{R \to ab}(p) = \Gamma^{(0)}_{R \to ab}
\left(\frac{p}{p_0}\right)^{2l+1} \left(\frac{p_0^2 +
\Lambda_{ab}^2}{p^2+\Lambda_{ab}^2}\right)^{l+1},
\end{equation}
where $p_0$ is a value of $p$ at the
resonance energy, $\Gamma^{(0)}_{R \to ab}=\Gamma_{R \to ab}(p_0)$ and $\Lambda_{ab}$ is a high-momentum cutoff parameter.
The same energy dependence was assumed for the total width of the $\mathcal{D}_{12}$ dibaryon $\Gamma_{D_{12}}$, however, its form has a little impact on the results presented below. In fact, we obtained very similar results with the constant $\Gamma_{D_{12}}$. The total width of the $\mathcal{D}_{03}$ dibaryon $\Gamma_{D_{03}}$ was assumed to be constant near the $\mathcal{D}_{03}$ resonance peak.

For the $pn \to \mathcal{D}_{03}$ vertex, we used the Gaussian
form factor, according to the dibaryon model for the $NN$
interaction~\cite{JPG01K,IJMP02K}. In this case, the
$\mathcal{D}_{03}$ decay width into the $np$ channel has the form
\begin{equation}
\Gamma_{D_{03} \to np}(p) \!=\! \Gamma_{D_{03} \to np}^{(0)}\!
\left(\frac{p}{p_0}\right)^5 \! {\rm
exp}\!\left(-\frac{p^2-p_0^2}{\Lambda_{pn}^2}\right). \label{gdnn}
\end{equation}

The cutoff parameters $\Lambda_{ab}$ were fixed by a
condition of a nearly constant width in the vicinity of the resonance position.
For the $\Delta \to \pi N$ vertices, we used the conventional value $\Lambda_{\pi N} = 0.16$ GeV/$c$~\cite{Bash17-ABC},
while we did not introduce any cutoff for the $\mathcal{D}_{03} \to \Delta\Delta$ vertex due to a compact size of the $\mathcal{D}_{03}$ resonance. So, we take the $\mathcal{D}_{03} \to \Delta\Delta$ vertex function simply as
\begin{equation}
\label{fdd}
F_{D_{03} \to
\Delta\Delta}(p_{\Delta\Delta}) = \sqrt{8\pi s}g_{\Delta\Delta},
\end{equation}
where $g_{\Delta\Delta}$ is a coupling constant. The parameters $\Gamma^{(0)}$ and $p_0$ (see Eq.~(\ref{G})) are not relevant here since $\mathcal{D}_{03}$ is not a resonance, but a bound state with regard to the $\Delta\Delta$ threshold.

We should note here that a soft cutoff parameter is justified for the $\Delta \to \pi N$ vertex form factor, since the pion can be considered as a point-like particle related to the nucleon, so, the soft cutoff means the peripheral character of the pion emission from the $\Delta$ isobar. The same is true for other meson-emission vertices. On the other hand, when we deal with the $\Delta\Delta$ component of the $\mathcal{D}_{03}$ resonance, two $\Delta$’s are almost fully overlapped, since $\mathcal{D}_{03}$ has the same size as the $\Delta$ (see the discussion in the beginning of this section). In such a case the $\mathcal{D}_{03} \to \Delta\Delta$ vertex form factor should be much harder. Since the precise cutoff value for this vertex is presently unknown, while it affects strongly the description of the low-mass part of the $M_{\pi\pi}$ spectrum~\cite{Adl11,Bash17-ABC}, we prefer to take the infinite cutoff here (which is equal to not introduce this form factor at all). In fact, we could either introduce no cutoff in the $\mathcal{D}_{03} \to np$ vertex, however, this vertex form factor is much less important near the $\mathcal{D}_{03}$ peak energy, since the $\mathcal{D}_{03}$ peak lies far from the $pn$ threshold. So, we keep the value $\Lambda_{pn}=0.35$ GeV/$c$, which follows from the dibaryon model of the $D$-wave $NN$ interaction~\cite{JPG01K}. Note that the cutoff value in the $\mathcal{D}_{03} \to \Delta\Delta$ vertex form factor should be larger, since there is no angular barrier in the $S$-wave $\Delta\Delta$ system. However, a two times smaller value of 0.16 GeV/$c$ was chosen in Refs.~\cite{Adl11,Bash17-ABC} to describe the low-mass $\pi\pi$ enhancement in the $pn \to d \pi^0\pi^0$ reaction by the $s$-channel $\Delta\Delta$ mechanism alone.

The parameters defined in Eqs.~(\ref{G}) and (\ref{gdnn}) as well as the masses and the total widths of the resonances used in the calculations below are listed in Tab.~\ref{tab1}. The parameter $g_{\Delta\Delta}$ defined in Eq.~(\ref{fdd}) will be discussed in the end of this section.

\begin{table}[!ht]
\flushleft {\caption{\label{tab1} Parameters of resonances $R$ and their decay
channels $R \to a+b$. For the parameter $p_0$, the given interval corresponds to all possible isospin channels.}}
\begin{tabular*}{\columnwidth}{@{\extracolsep{\fill}}cccccccc} \hline\hline $R$ & $M_R$ &
$\Gamma_R^{(0)}$ & $ab$ & $l$ &
$p_0$ & $\Gamma_{R\to ab}^{(0)}$ & $\Lambda_{ab}$ \\
& (MeV) & (MeV) & & & (MeV) & (MeV) & (GeV) \\
\hline
& & & $np$ & 2 & $730$ & $9$ & $0.35$ \\ 
$\mathcal{D}_{03}$ & $2376$ & $77$ & $\sigma d$ & $2$ & $350$ & $2$ & $0.18$ \\
& & & $\pi \mathcal{D}_{12}$ & $1$ & $173$--$176$ & $31$ & $0.12$
\\
$\mathcal{D}_{12}$ & $2150$ & $110$ & $\pi d$ & $1$ & $221$--$223$ & $33$ &
$0.15$
\\ 
$\Delta$ & $1232$ & $117$ & $\pi N$ & $1$ & $226$--$229$ & $117$ &
$0.16$
\\ 
$\sigma$ & $303$ & $126$ & $\pi\pi$ & $0$ & $72$--$80$ & 126 & $0.09$
\\
\\ \hline\hline
\end{tabular*}
\end{table}


The total mass and width of the $\mathcal{D}_{03}$ resonance have been fixed in accordance with the experimental data~\cite{Adl11,Adl14-el} and fine tuned to fit the total $pn \to d \pi^0\pi^0$ cross section from Ref.~\cite{Adl13-iso} in the range $\sqrt{s}=2.36$--$2.40$ GeV (see Sec.~\ref{sec-ed} and Appendix for the normalization issue). In turn, the total mass and width of the $\mathcal{D}_{12}$ resonance have been fixed as in Ref.~\cite{PRC2013} to be consistent with the available experimental and PWA data (see, e.g., Ref.~\cite{Hoshizaki93}), and they are also consistent with the Faddeev calculations of the $\pi NN$ system~\cite{Gal14} and the model calculations of the $pp \to d \pi^+$ reaction~\cite{NPA2016}. On the other hand, the $\sigma$-meson mass and width have been found from the fit to the ABC peak obtained in Ref.~\cite{Adl13-iso}. Then, the partial decay width of the $\sigma$-meson into $\pi^0\pi^0$ and $\pi^+\pi^-$ channels are found from the total width by isospin relations, taking into account the kinematic consequences of the 5-MeV mass difference between the charged and neutral pions (see Sec.~\ref{sec-sig}). In the initial version of the model~\cite{PRC2013,NPA2016}, we neglected the pion mass difference, however, it should be taken explicitly into account to give the quantitative predictions for both reactions $pn
\to d \pi^0\pi^0$ and $pn \to d \pi^+\pi^-$ in the near-threshold region. As will be shown in two next sections, this refinement of the initial model gives the most visible consequences for the $\sigma$-meson production mechanism, while it is practically negligible (except for the phase space difference) for the $\mathcal{D}_{12}$ and $\Delta\Delta$ excitation mechanisms. Hence, the partial widths of the $\mathcal{D}_{12}$ (or the $\Delta$) decay into different $\pi d$ (or $\pi N$) channels can safely be fixed by isospin relations. For the $\mathcal{D}_{12} \to \pi d$ BR, we adopted the value of 30\%, which follows from the SAID PWA~\cite{Arndt94-pid}. Thus, we obtained the value of $\Gamma^{(0)}_{D_{12}\to \pi d}=33$ MeV for the total $\mathcal{D}_{12}$ width of 110 MeV. At this point, it is important to realise that the $\mathcal{D}_{12}$ resonance has been treated in the literature in two different ways. If it is treated as a pure dibaryon state as, e.g., in Refs.~\cite{NPA2016,Simonov79}, it should be supplemented by the $t$-channel $N\Delta$ excitation and other less important background processes in the $^1D_2$ $pp$ (or $^3P_2$ $\pi d$) partial channel. In this case, one obtains the BR for the $\mathcal{D}_{12} \to NN$ and $\mathcal{D}_{12} \to \pi d$ decays of about 10\% or less, as we found from the fit of the $pp \to d\pi^+$ cross section in the ${}^1D_2p$ partial wave~\cite{NPA2016}. Alternatively, if the $\mathcal{D}_{12}$ state is supposed to saturate the relevant channel, as in the model~\cite{Gal17,Gal14}, the intermediate $N\Delta$ state is treated as its component, and its BR can be read off the Argand diagrams obtained in PWA. In this case, one obtains from the SAID PWA the BR for the $\mathcal{D}_{12} \to NN$ decay of about 16--18\%~\cite{Arndt07,Arndt87} and for the $\mathcal{D}_{12} \to \pi d$ decay of about 30\%~\cite{Arndt94-pid}. The same value for the latter BR was found in the Gatchina PWA~\cite{Strak83,Strak91} (though the $\mathcal{D}_{12} \to NN$ BR was found there to be of about 10\% only). When we consider here the decay $\mathcal{D}_{03} \to \mathcal{D}_{12} + \pi$, we effectively take into account the $\pi +N +\Delta$ intermediate state along with the $\mathcal{D}_{12}+\pi$ one, thus, we should take the $\mathcal{D}_{12} \to \pi d$ BR from the PWA data.

We complete our model description with the discussion of the parameters $\Gamma^{(0)}$ for the $\mathcal{D}_{03}$ production and decay. In fact, by fitting the experimental $M_{\pi\pi}$ distributions, we can find only the products of the incoming and outgoing partial widths. So, we need to fix some of them from the independent sources. Thus, we fixed the incoming width $\Gamma^{(0)}_{D_{03}\to pn}$ to be 12\% of the total $\mathcal{D}_{03}$ width, according to the experimental data on $np$ elastic scattering (see Ref.~\cite{Bash15-BR}). Then the value of $\Gamma^{(0)}_{D_{03}\to d\sigma}$ was found from the fit of the respective amplitude to the $M_{\pi\pi}$ spectra in the low-mass region, provided the summed contribution of the $\mathcal{D}_{12}\pi$ and $\Delta\Delta$ mechanisms was found from the fit to the $M_{\pi\pi}$ spectra in the high-mass region. However, since both these mechanisms give very similar results for the double-pion production cross sections, the question arises about their relative weight in the $\mathcal{D}_{03}$ decay.
 The contribution of the $\mathcal{D}_{03} \to \mathcal{D}_{12}\pi$ decay mode to the $pn \to d (\pi\pi)_0$ cross sections can be restricted by the experimental data as follows. The $\mathcal{D}_{03} \to d \pi\pi$ decay branch is about 37\% of the total $\mathcal{D}_{03}$ width~\cite{Bash15-BR}. The contributions of the $\mathcal{D}_{03} \to \mathcal{D}_{12}\pi$ mode to the $\mathcal{D}_{03}$ decay into $d\pi\pi$ and $NN\pi$ final states are related as the BR for the $d\pi$ and $NN$ decays of the $\mathcal{D}_{12}$ resonance. From the SAID PWA~\cite{Arndt94-pid,Arndt07,Arndt87}, these BR are found to be related approximately as 2:1. Since the $\mathcal{D}_{03} \to NN\pi$ decay, if it takes place at all, can proceed predominantly through the $\mathcal{D}_{12}+\pi$ intermediate state, the upper limit for the $\mathcal{D}_{03} \to NN\pi$ BR of 5\% found in Ref.~\cite{Adl17-NNpi} means the upper limit for the $\mathcal{D}_{03} \to \mathcal{D}_{12}\pi \to d\pi\pi$ mode to be about 10\% of the $\mathcal{D}_{03}$ total width, i.e., about 25\% of the $\mathcal{D}_{03} \to d \pi\pi$ partial width.\footnote{Note that for the Gatchina PWA~\cite{Strak83,Strak91}, this upper limit would be about 40\%, since the BR for the $d\pi$ and $NN$ decays of the $\mathcal{D}_{12}$ resonance are related there as 3:1.} We adopted just this value here (i.e., 3 times smaller than in the initial model~\cite{PRC2013}), which leads to the $\Gamma^{(0)}_{D_{03}\to D_{12}\pi}$ value listed in Tab.~\ref{tab1}.

Then, from the fit of the high-mass $M_{\pi\pi}$ spectra, we obtained the 20\% contribution of the $\mathcal{D}_{03}\to \Delta\Delta$ mode to the $pn \to d \pi^0\pi^0$ cross section at $\sqrt{s}=2.38$ GeV, which corresponds to the coupling constant $g_{\Delta\Delta}=1.23$ (see Eq.~(\ref{fdd})). This value, as well as the values for the parameters $\Gamma^{(0)}_{D_{03}\to d\sigma}$ and $\Gamma^{(0)}_{D_{03}\to D_{12}\pi}$ given in Tab.~\ref{tab1}, correspond to the experimental total cross section $\sigma(pn \to d\pi^0\pi^0)= 0.255$~mb at $\sqrt{s}=2.38$ GeV~\cite{Adl13-iso}. Thus, we have four adjustable parameters in the model: the mass and width of the $\sigma$-meson, the partial width $\Gamma^{(0)}_{D_{03}\to d\sigma}$, and the coupling constant $g_{\Delta\Delta}$.

\section{Neutral and charged dipion production via the intermediate $\bm{\mathcal{D}_{12}}$ and $\bm{\Delta\Delta}$ excitation}
\label{sec-dd}

In this section, we study the two-pion invariant-mass spectra at $\sqrt{s}=2.38$ GeV which result from the dominant mechanism of the reactions $pn \to d (\pi\pi)_0$ in the region of the $d^*(2380)$ excitation. Such a mechanism can include either intermediate $\mathcal{D}_{12}(2150)$ or $\Delta\Delta$ excitation, as proposed in Refs.~\cite{PRC2013} and~\cite{Adl11},
respectively. Here, we present the calculations for each of these mechanisms separately with regard to their impact on the observed difference between the $\pi^0\pi^0$ and $\pi^+\pi^-$ production cross sections in $pn$ collisions. In the next section, we will combine both above mechanisms with the intermediate $\sigma$-meson excitation and analyze the predictions of the full model.

From isospin conservation, one expects for the total double-pion production cross
sections in various isospin channels:
\begin{equation}
\label{isosig}
\sigma(pn \to d \pi^+\pi^-) = 2 \sigma(pn \to d \pi^0\pi^0)+
\frac{1}{2}\sigma(pp \to d \pi^+\pi^0).
\end{equation}

We consider here the isoscalar dipion production, which is connected to the formation of the $\mathcal{D}_{03}$ resonance. Hence, we compare our calculations for the cross sections in the reaction $pn \to d\pi^+\pi^-$ to the experimental data on $\sigma(pn \to d
(\pi^+\pi^-)_0) = \sigma(pn \to d \pi^+\pi^-) - \textstyle{\frac{1}{2}}\sigma(pp \to d \pi^+\pi^0)$. Though the different channels have different thresholds in the $\pi\pi$ invariant mass, the above subtraction can be performed safely for the data~\cite{Adl13-iso}, since the data for different isospin channels have been averaged over the same 10-MeV intervals in $M_{\pi\pi}$. In any case, the isovector $\pi\pi$ channel is suppressed near threshold due to the Pauli principle and therefore, its contribution to the $\pi^+\pi^-$ production cross section is significant only at rather high values of $M_{\pi\pi}$. In turn, according to Eq.~(\ref{isosig}), the cross section $\sigma(pn \to d (\pi^+\pi^-)_0)$ should be compared with $2\sigma(pn \to d \pi^0\pi^0)$ to explore the isospin symmetry breaking in isoscalar dipion production.

\begin{figure}[!ht]
\begin{center}
\resizebox{1.0\columnwidth}{!}{\includegraphics{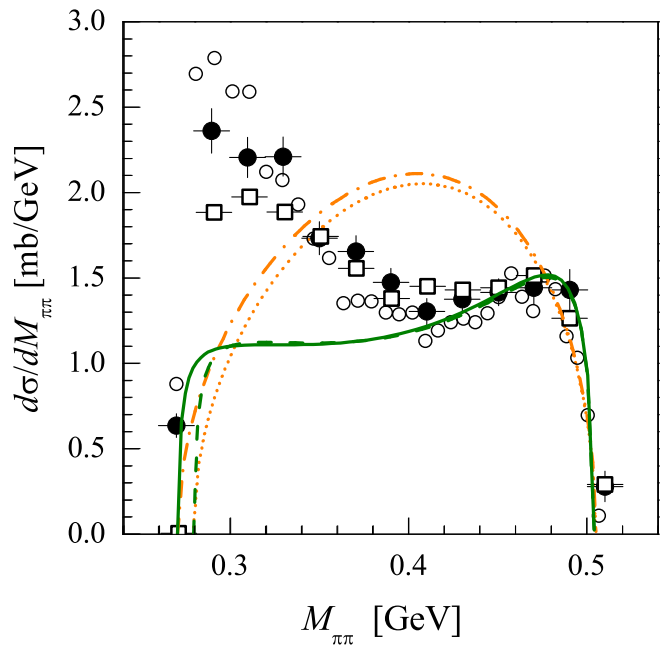}}
\end{center}
\caption{(Color online) The $\pi\pi$ invariant-mass distributions in the reactions $pn \to d \pi^0\pi^0$ (multiplied by 2, solid line) and $pn \to d (\pi^+\pi^-)_0$ (dashed line) at $\sqrt{s}=2.38$ GeV resulted from the $\mathcal{D}_{03}\to \mathcal{D}_{12}\pi$ decay in the intermediate state. The theoretical calculations are compared to the experimental data on $2d\sigma/dM_{\pi^0\pi^0}$ (filled circles) and $d\sigma/dM_{\pi^+\pi^-}-\frac{1}{2}d\sigma/dM_{\pi^+\pi^0}$ (open squares) from Ref.~\cite{Adl13-iso}, as well as the data on $2d\sigma/dM_{\pi^0\pi^0}$ from Ref.~\cite{Adl11} (open circles). The latter data have been multiplied by 0.45 (see Appendix). The model parameters are those listed in Tab.~\ref{tab1}, except for the parameter $\Gamma^{(0)}_{{D}_{03}\to {D}_{12}\pi}$, which has been adjusted to reproduce the experimental data at high invariant masses. Also shown are the pure phase-space distributions for $\pi^0\pi^0$ (dash-dotted line) and $\pi^+\pi^-$ (dotted line) production normalized to the respective total cross sections.
} \label{fig-d12}
\end{figure}

In Fig.~\ref{fig-d12} we show the $\pi\pi$ invariant-mass distributions in the reactions $pn \to d\pi^0\pi^0$ and $pn \to d\pi^+\pi^-$ calculated for the $\mathcal{D}_{12}$ production mechanism. The calculations for the neutral and charged dipion production channels were performed with the same model parameters (including the overall normalization) but different pion masses. As was found in Ref.~\cite{PRC2013}, the $\mathcal{D}_{12}$ excitation mechanism alone can give a reasonable (at least qualitative) description of the high-$M_{\pi\pi}$ data but lacks any low-$M_{\pi\pi}$ enhancement (the ABC effect). As is clearly seen from Fig.~\ref{fig-d12}, this mechanism gives a shift of the low-$M_{\pi\pi}$ distribution due to the phase space reduction for the charged dipion channel, but does not lead to any suppression of the charged dipion production cross section at low invariant masses.

\begin{figure}[!ht]
\begin{center}
\resizebox{1.0\columnwidth}{!}{\includegraphics{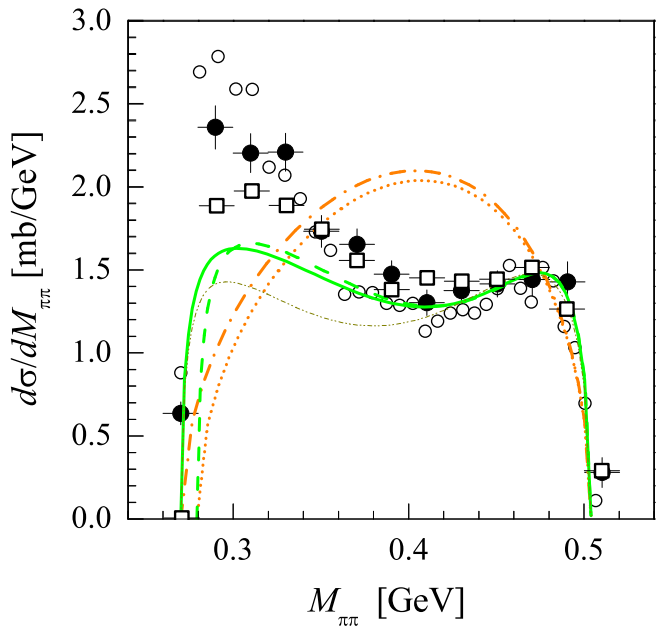}}
\end{center}
\caption{(Color online) The same as in Fig.~\ref{fig-d12}, but for the $\mathcal{D}_{03}\to \Delta\Delta$ decay in the intermediate state. The coupling constant $g_{\Delta\Delta}$ has been adjusted to reproduce the experimental data at high invariant masses. The thin dash-dot-dotted line shows the result of the calculation for $\pi^0\pi^0$ production without account for the nucleon recoil in the $\Delta \to \pi N$ decay (as in Ref.~\cite{Bash17-ABC}).
} \label{fig-dd}
\end{figure}

The $\pi\pi$ invariant-mass distributions for the $s$-channel $\Delta\Delta$ excitation in the intermediate state are shown in Fig.~\ref{fig-dd}. Though a moderate low-mass enhancement in the $M_{\pi\pi}$ spectra is present here, it is not sufficient to reproduce the observed strength of the ABC effect. In fact, the low-mass enhancement seen in Fig.~\ref{fig-dd} comes mainly from the nucleon recoil in the $\Delta\to N\pi$ vertices, which was neglected in the calculations~\cite{Adl11,Bash17-ABC}. We also found sensitivity of the $\Delta\Delta$ mechanism contribution to the cutoff parameter in the $\Delta\to \pi N$ vertices. For instance, if we take $\Lambda_{\pi N}=0.3$~GeV/$c$~\cite{NPA2016} instead of the conventionally used value (for the on-shell pion) of 0.16~GeV/$c$~\cite{Bash17-ABC}, the low-mass enhancement will be ca. 10\% higher, but still too low compared to the experimental ABC peak, especially when comparison is made with the more precise data~\cite{Adl11} (shown in Fig.~\ref{fig-dd} by open circles). These data can be described well only if the soft form factor in the $\mathcal{D}_{03}\to \Delta\Delta$ vertex is introduced with the cutoff value $\Lambda_{\Delta\Delta}=0.15$--$0.2$~GeV/$c$~\cite{Adl11}. Such a soft cutoff is appropriate for a loosely bound (deuteron-like) object, but is hardly compatible with the compact size of the $\mathcal{D}_{03}$ state (r.m.s. of about 0.8 fm), its high binding energy (ca. 80 MeV) in the $\Delta\Delta$ channel and its narrow width (see also the discussion in Sec.~\ref{sec-form}). Furthermore, the cutoff parameter $\Lambda_{\Delta\Delta}$ should be even smaller (ca. 0.07~GeV/$c$) to reproduce the ABC enhancement in $dd$ collisions~\cite{Adl12-dd}.

As is seen from Fig.~\ref{fig-dd}, the $s$-channel $\Delta\Delta$ mechanism does not give any suppression of the ABC peak for the charged dipion production. Similarly to the $\mathcal{D}_{12}$ excitation mechanism, it gives only a shift of the low-mass distribution in the $\pi^+\pi^-$ channel by about 10 MeV, which comes from the pion mass difference and the corresponding phase space reduction. In fact, the low-mass enhancement in the $\pi^+\pi^-$ channel turns out to be even a bit higher than that in the $\pi^0\pi^0$ channel. Inclusion of the above $\mathcal{D}_{03}\to \Delta\Delta$ vertex form factor cannot help here, since the form factor should be \textit{the same} for the neutral and charged dipion production channels.


Thus, we conclude that none of the mechanisms leading to production of two uncorrelated pions (emitted from two different resonances), i.e., with the isovector $\mathcal{D}_{12}$ dibaryon or two $\Delta$ isobars in the intermediate state, gives the observed suppression of the charged dipion production cross section at low $\pi\pi$ invariant masses compared to that for neutral dipion production. Both these mechanisms give only a shift of the low-$M_{\pi\pi}$ distribution due to a 10-MeV shift of the dipion production threshold in the charged channel. This is not surprising since the resonances ($\mathcal{D}_{03}$, $\mathcal{D}_{12}$, and $\Delta$), which produce the pions in the above mechanisms, are located far from the respective pion-production thresholds, and thus their decay is weakly sensitive to the small mass difference between the charged and neutral pions. Hence, one should seek for another source of the ABC peak itself, as well as for the reduction of its strength in the $\pi^+\pi^-$ production channel. In the next section, the intermediate scalar $\sigma$-meson production will be considered as a possible candidate.

 We should emphasise here that we used the double-pion production amplitude, which is explicitly isospin-violating due to the pion mass difference. Thus, we used the charged pion mass in the amplitude of charged dipion production and the neutral pion mass in the amplitude of neutral dipion production. However, in Ref.~\cite{Adl13-iso} the amplitude of the $s$-channel $\Delta\Delta$ mechanism was treated differently. This amplitude was made explicitly isospin-symmetric by taking equal pion masses in the amplitudes for two above dipion production channels (but different pion masses in the phase-space factors). In fact, the isospin symmetry is often assumed for the amplitudes of hadronic processes when the accurate dynamical treatment is not available (see, e.g.,~\cite{Wilkin00}). This way the observed strong isospin symmetry violation in the near-threshold region can be explained by solely the phase-space difference for neutral and charged dipion production (see Fig.~2 in Ref.~\cite{Adl13-iso}). Hence, the calculation of the $\Delta\Delta$ mechanism in Ref.~\cite{Adl13-iso} (supplemented by the soft $\mathcal{D}_{03} \to \Delta\Delta$ form factor) turned out to be in qualitative agreement with the data~\cite{Adl13-iso}, which indeed show at energies $\sqrt{s}\simeq 2.38$ GeV the isospin symmetry violation close to that contained just in the phase-space factors. On the contrary, in our model, we do not make the amplitude isospin-symmetric by hand, but incorporate the real pion masses in it.\footnote{We believe that the mass difference between final pions has the biggest impact on the isospin symmetry violation near the two-pion threshold, both in the amplitude and phase space, while the mass difference between the intermediate $\Delta$ isobars (or different charge states of the $\mathcal{D}_{12}$ dibaryon) can be neglected.} Since our amplitude depends on the $\pi\pi$ relative momentum, which rises from zero at threshold in both dipion channels, both amplitudes fall rapidly from (almost) the same values at the respective thresholds with the rising $M_{\pi\pi}$ and approach very close values at high $M_{\pi\pi}$. Thus, the amplitude for the neutral dipion production occurs to be much lower than that for the charged dipion production at the $\pi^+\pi^-$ threshold. This isospin-violating behaviour of the amplitudes in the near-threshold region is compensated by the opposite behaviour of the phase-space factors thus leading to the result plotted in Fig.~\ref{fig-dd}.
Hence, we need an additional dynamical mechanism which would partially restore the isospin symmetry of the {\em total} $2\pi$-production amplitude and improve agreement with the data.

\section{Inclusion of the intermediate $\bm{\sigma}$-meson production}
\label{sec-sig}

We have shown in Ref.~\cite{PRC2013} that the ABC effect in the reaction $pn \to d \pi^0\pi^0$ can be explained by the intermediate $\sigma$-meson excitation mechanism, i.e., $pn \to \mathcal{D}_{03} \to d+\sigma \to d + \pi^0\pi^0$. If we add the respective amplitude coherently to the amplitude of the intermediate $\mathcal{D}_{12}$ excitation, the sum of these two amplitudes gives a pronounced low-mass enhancement in the $M_{\pi\pi}$ spectrum, provided the $\sigma$ mass and width are shifted downwards from their free-space values (listed by PDG~\cite{PDG20}) due to the partial chiral symmetry restoration in the $\mathcal{D}_{03}$ dibaryon~\cite{PRC2013}. So, the observed strength and position of the ABC enhancement can be reproduced with $m_{\sigma} \simeq 300$ MeV and $\Gamma_{\sigma} \simeq 100$ MeV. We note in passing that according to the well-established point of view (see, e.g., Refs. \cite{Glozman00,Glozman07}), the partial chiral symmetry restoration, which leads to the shift of the $\sigma$-meson mass towards the $2\pi$ threshold, can occur in highly excited hadrons due to decoupling of the valence quarks from the QCD condensates. In this respect, the above dibaryon state $\mathcal{D}_{03}$ having the mass $M_{D_{03}}\simeq 2.38$ GeV, i.e., 500 MeV above the $pn$ threshold, can be considered as a highly excited hadronic state.

In the work~\cite{PRC2013} we did not take into account the difference between the neutral and charged pions when considering the $\sigma$-meson decay. When the pion mass difference is taken into account, the total decay width of
the $\sigma$ meson can no longer be taken as
$\Gamma_{\sigma}(M_{\pi\pi})=3\Gamma_{\sigma \to
\pi^0\pi^0}(M_{\pi\pi})$, but rather should be
\begin{equation}
\Gamma_{\sigma}(M_{\pi\pi}) = \Gamma^{(0)}_{\sigma n}
\frac{k_n}{k_{0n}}\frac{k_{0n}^2+\Lambda_{\pi\pi}^2}{k_n^2+\Lambda_{\pi\pi}^2}
+ \Gamma^{(0)}_{\sigma c}
\frac{k_c}{k_{0c}}\frac{k_{0c}^2+\Lambda_{\pi\pi}^2}{k_c^2+\Lambda_{\pi\pi}^2},
\label{gst}
\end{equation}
where $k_n^2 = M_{\pi\pi}^2/4-m_{\pi^0}^2$ and $k_c^2 =
(M_{\pi\pi}^2/4-m_{\pi^+}^2)\theta(M_{\pi\pi}-2m_{\pi^+})$ are the moduli squared of the relative momenta of two neutral and charged pions, respectively, as functions of $M_{\pi\pi}$. We use here the same cutoff parameter
$\Lambda_{\pi\pi}$ for the charged and neutral pions. At the resonance energy, one has $M_{\pi\pi} = m_{\sigma}$,
$k_{n}=k_{0n}$, $k_{c}=k_{0c}$, and the $\sigma$ decay widths into the neutral and charged two-pion channels are $\Gamma^{(0)}_{\sigma n}$ and $\Gamma^{(0)}_{\sigma c}$, respectively. In case of isospin conservation, one would get $k_{n}=k_{c}$ and $\Gamma^{(0)}_{\sigma n}=\Gamma^{(0)}_{\sigma c}/2 = \Gamma^{(0)}_{\sigma}/3$.

When the isospin symmetry violation is considered explicitly, it is convenient to introduce the coupling constants $g^2_{\sigma n} = \Gamma^{(0)}_{\sigma n}/k_{0n}$ and $g^2_{\sigma c} = \Gamma^{(0)}_{\sigma c}/k_{0c}$. This allows to separate the basic isospin dependence of the $\sigma$ partial decay widths due to the different $\pi\pi$ relative momenta at $M_{\pi\pi} = m_{\sigma}$ for the different pion masses.
If to assume, as usual,
\begin{equation}
\label{gnc0}
g^2_{\sigma n} = g^2_{\sigma c}/2 = g^2_{\sigma},
\end{equation}
then the coupling constant $g_{\sigma}$ is uniquely related to the
total width $\Gamma_{\sigma}$ at the resonance point $M_{\pi\pi} = m_{\sigma}$.
In this case, the difference between the $\sigma$ partial decay widths into the neutral and charged dipions (aside from a factor of $1/2$) is governed by the kinematical difference between the momenta $k_n$ and $k_c$. Below we will also consider the possible
dynamical origin of the difference between the $\sigma$ partial widths by introducing an additional
parameter $\alpha$, so that,
\begin{equation}
\label{gnc}
g^2_{\sigma n} = g^2_{\sigma}(1+\alpha), \quad g^2_{\sigma c} =
g^2_{\sigma}(2-\alpha).
\end{equation}

\begin{figure}[!ht]
\begin{center}
\resizebox{1.0\columnwidth}{!}{\includegraphics{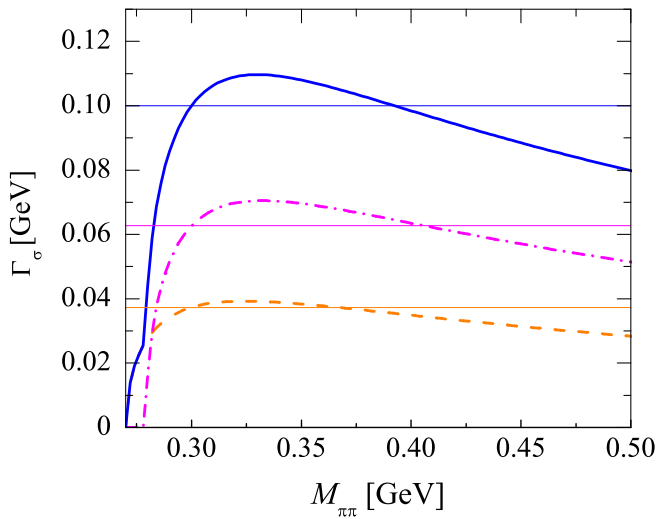}}
\end{center}
\caption{(Color online) The partial decay widths $\Gamma_{\sigma \to
\pi^0\pi^0}$ (dashed line) and $\Gamma_{\sigma \to \pi^+\pi^-}$ (dash-dotted
line), as well as the total width $\Gamma_{\sigma}$ (solid line)
as functions of the $\pi\pi$ invariant mass, according to
Eq.~(\ref{gst}), for $m_{\sigma}=300$ MeV and $\Gamma_{\sigma}(m_{\sigma}) = 100$
MeV~\cite{PRC2013}. The constant
values of the widths at $M_{\pi\pi} = m_{\sigma}$ are shown by
thin solid lines.} \label{fig-gs}
\end{figure}

The $M_{\pi\pi}$ dependence of the total and partial $\sigma$ widths for $m_{\sigma}=300$ MeV, $\Gamma_{\sigma} = 100$ MeV~\cite{PRC2013}, and $\alpha=0$ is plotted in Fig.~\ref{fig-gs}.

\begin{figure*}[!ht]
\begin{center}
\resizebox{1.0\textwidth}{!}{\includegraphics{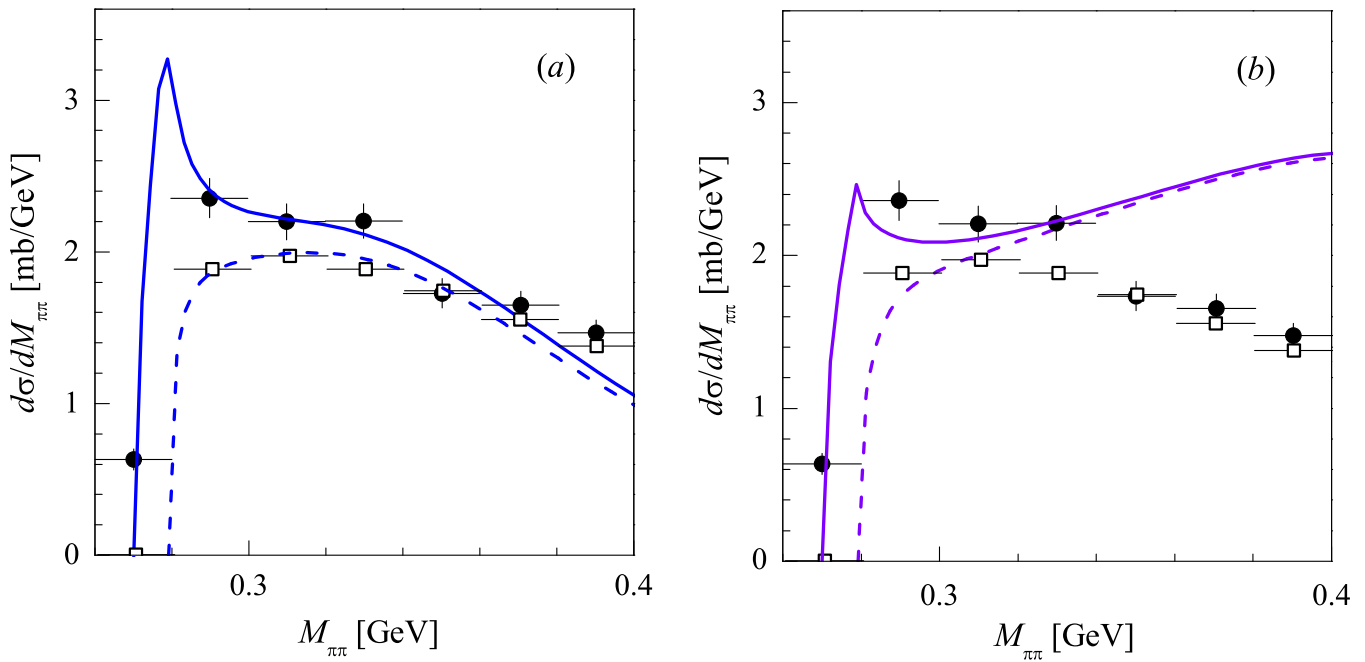}}
\end{center}
\caption{(Color online) ($a$) The $\pi\pi$ invariant-mass distributions in the reactions $pn \to d \pi^0\pi^0$ (multiplied by 2, solid line) and $pn \to d (\pi^+\pi^-)_0$ (dashed line) at $\sqrt{s}=2.38$ GeV resulted from the $\mathcal{D}_{03} \to d +\sigma$ decay in the intermediate state. The theoretical calculations are compared to the experimental data on $2d\sigma/dM_{\pi^0\pi^0}$ (filled circles) and $d\sigma/dM_{\pi^+\pi^-}-\frac{1}{2}d\sigma/dM_{\pi^+\pi^0}$ (open squares) taken from Ref.~\cite{Adl13-iso}. The $\sigma$-meson mass and width and the $\mathcal{D}_{03} \to d +\sigma$ decay width have been adjusted to reproduce the low-$M_{\pi\pi}$ data. ($b$) The same as ($a$), but for the fixed values $m_{\sigma}=\Gamma_{\sigma}=500$ MeV.} \label{fig-sig}
\end{figure*}

The above structure of the total $\sigma$ width leads to a singularity in the $\pi^0\pi^0$ production cross section at the $\pi^+\pi^-$  threshold, which is absent in the $\pi^+\pi^-$ production cross section. The $M_{\pi\pi}$ spectra resulted from the intermediate $\sigma$ excitation mechanism with the width parametrization~(\ref{gst}) are shown in Fig.~\ref{fig-sig}. In Fig.~\ref{fig-sig}$a$ we have fitted the $\sigma$ mass and width and the overall normalization of the cross section (which is defined by the $\mathcal{D}_{03} \to d+\sigma$ decay width) to reproduce the experimental distributions~\cite{Adl13-iso} at low $M_{\pi\pi}$. We obtained the values of $m_{\sigma}=322$ MeV and $\Gamma_{\sigma}=158$ MeV, which are somewhat higher than the values found in Ref.~\cite{PRC2013}. This difference is due to the broader ABC peak in the data~\cite{Adl13-iso} compared to that in the previous data~\cite{Adl11} and also due to inclusion of the ``background'' $\mathcal{D}_{12}$ contribution in the fit~\cite{PRC2013}.\footnote{
Both data sets~\cite{Adl11,Adl13-iso} are shown in Figs.~\ref{fig-d12} and~\ref{fig-dd}, but we do not plot the data~\cite{Adl11} on the $M_{\pi\pi}$ distribution for $\pi^0\pi^0$ production in the next Figures. We focus here on the isospin symmetry violation, which can be traced by analyzing the data on dipion production in different isospin channels {\em measured in the same experiment}. The data~\cite{Adl11}, though being more precise, differ significantly from the data~\cite{Adl13-iso} at low $M_{\pi\pi}$, therefore, we do not include the data~\cite{Adl11} in our present analysis.} It is seen from Fig.~\ref{fig-sig}$a$ that the striking difference between the neutral and charged dipion production cross sections in the isoscalar channel, which results from the structure of the $\sigma$ decay width~(\ref{gst}) shown in Fig.~\ref{fig-gs}, is in agreement with the experimental data. In Fig.~\ref{fig-sig}$b$ we also show the $M_{\pi\pi}$ spectra obtained for the values $m_{\sigma}=\Gamma_{\sigma}=500$ MeV consistent with those listed in the PDG tables~\cite{PDG20}. In this case the cusp in the $\pi^0\pi^0$ invariant-mass distribution at the $\pi^+\pi^-$ production threshold is also visible, however the shape of the distribution differs strongly from the experimental one. In fact, the calculated distribution rises up to the nominal $\sigma$ mass, while the experimental one decreases after the low-mass peak. So, the data on double-pion production favor the lower mass and width of the $\sigma$ meson.

It is worth emphasising that the singular behavior of the production cross section in the given channel at the threshold of another channel (with a higher threshold) is characteristic for excitation of an intermediate resonance $R$, which can decay into both channels and is located near their thresholds, i.e., when the resonance mass and width satisfy the relation $M_R-M_{\rm thr} < \Gamma_R/2$. This is related to the fact that the detailed structure of the decay width is important mainly near the resonance position. This condition is fulfilled for the narrow near-threshold $\sigma$ meson with the mass $m_{\sigma} \sim 300$ MeV and the width $\Gamma_{\sigma} \sim 100$ MeV, as well as for the broad $\sigma$-meson with the mass and width $m_{\sigma} \sim \Gamma_{\sigma} \sim 500$ MeV. However, it is not the case for the $\Delta$ or $\mathcal{D}_{12}$ resonances with respect to their single-pion decays. Both these resonances are located rather far from the pion production thresholds, hence the production cross sections via these resonances are only slightly affected by the small difference between the neutral and charged pion production thresholds. That is why, when taking into account explicitly the pion mass difference in the total width parametrization for the $\Delta$ or $\mathcal{D}_{12}$ resonances, we do not find any significant difference between the neutral and charged dipion production cross sections, except for some shift of the low-mass peak in the $M_{\pi\pi}$ spectrum due to the phase-space reduction for the charged dipions (see Sec.~\ref{sec-dd}).

\begin{figure*}[!ht]
\begin{center}
\resizebox{1.0\textwidth}{!}{\includegraphics{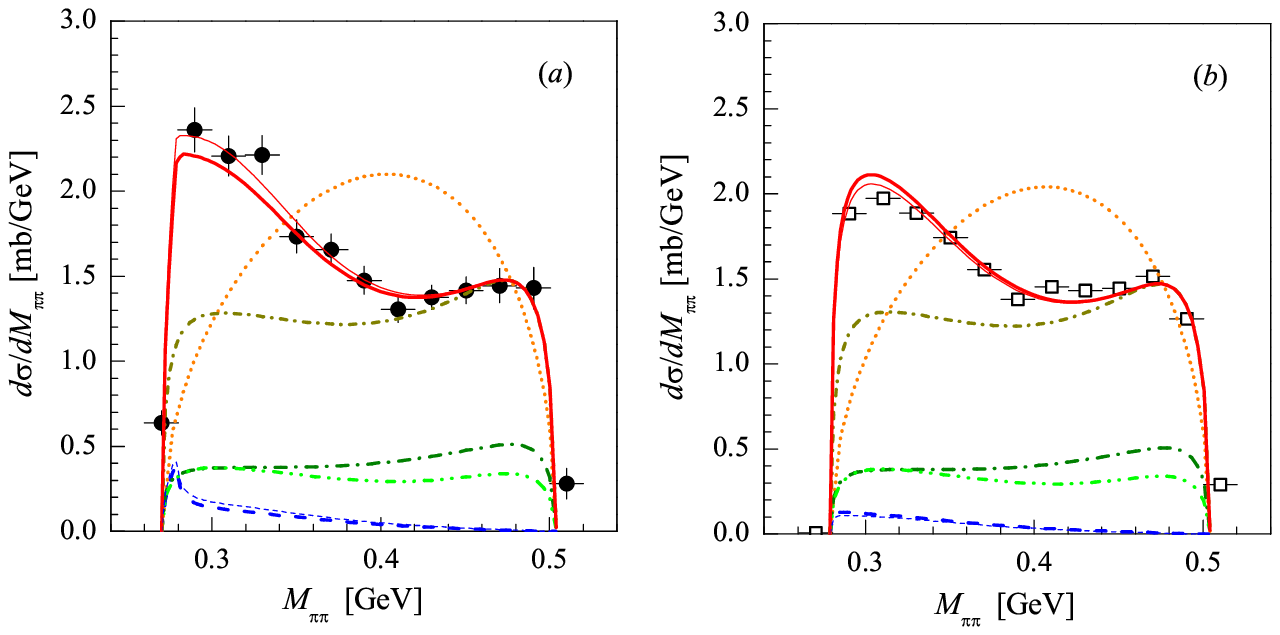}}
\end{center}
\caption{(Color online) The $\pi\pi$ invariant-mass distributions in the reactions $(a)$ $pn \to d \pi^0\pi^0$ (multiplied by 2) and $(b)$ $pn \to d (\pi^+\pi^-)_0$ at $\sqrt{s}=2.38$ GeV calculated with the model parameters from Tab.~\ref{tab1}. Shown are the distributions resulted from the $\mathcal{D}_{03} \to \mathcal{D}_{12}+\pi$ decay (dash-dotted lines), the $\mathcal{D}_{03} \to \Delta+\Delta$ decay (dash-dot-dotted lines), the $\mathcal{D}_{03}\to d+\sigma$ decay (dashed lines), and the coherent sum of these three $\mathcal{D}_{03}$ decay routes (solid lines). Upper dash-dotted lines (with short dashes) show the summed contribution of the $\mathcal{D}_{12}+\pi$ and $\Delta+\Delta$ excitation mechanisms. Dotted lines correspond to the pure phase-space distributions. Thin dashed and solid lines correspond to the $\sigma$-excitation mechanism and the total distributions with $\alpha=0.23$ (see Eq.~(\ref{gnc})). The theoretical calculations are compared to the experimental data on $2d\sigma/dM_{\pi^0\pi^0}$ (filled circles) and $d\sigma/dM_{\pi^+\pi^-}-\frac{1}{2}d\sigma/dM_{\pi^+\pi^0}$ (open squares) taken from Ref.~\cite{Adl13-iso}.} \label{fig-tot}
\end{figure*}

Thus, we have shown that the near-threshold $\sigma$-meson production can explain the observed suppression of the ABC enhancement in the $\pi^+\pi^-$ channel. However, when the $\sigma$-excitation amplitude is added coherently to the $\mathcal{D}_{12}$ or $\Delta\Delta$ production amplitude to reproduce also the high-mass part of the $M_{\pi\pi}$ spectrum, the contribution of the $\sigma$-excitation mechanism should be decreased in comparison to that shown in Fig.~\ref{fig-sig}$a$. Then its influence on the summed $M_{\pi\pi}$ distribution gets reduced. In Fig.~\ref{fig-tot} we show the $M_{\pi\pi}$ distribution resulted from the coherent sum of the $\sigma$, $\mathcal{D}_{12}$, and $\Delta\Delta$ excitation mechanisms, as well as their individual contributions. The integrated contributions of these mechanisms to the $pn \to d \pi^0\pi^0$ cross section are about 4\%, 25\%, and 20\%, respectively. For the better visibility of all curves, we depicted the distribution for $\pi^0\pi^0$ and $\pi^+\pi^-$ production in two separate figures. The resonance parameters used in calculations for this version of the model are listed in Tab.~\ref{tab1}.\footnote{For comparison with the data~\cite{Adl13-iso} on the differential $M_{\pi\pi}$ distributions, presented in Figs.~\ref{fig-tot}--\ref{fig-ds3e}, we decreased the absolute normalization of our cross sections by 1.32 --- see Appendix.} In particular, we obtained the values of $m_{\sigma}=303$ and $\Gamma_{\sigma}=126$ MeV (denoted in Tab.~\ref{tab1} as $M_{\sigma}$ and $\Gamma^{(0)}_{\sigma}$). In should be noted that the above refinement of the total $\sigma$ width by inclusion of the pion mass difference (see Eq.~(\ref{gst})) and adding the $s$-channel $\Delta\Delta$ mechanism also lead to some modification in the description of the $M_{\pi\pi}$ distribution in the reaction $pn \to d \pi^0\pi^0$~\cite{Adl11} published in Refs.~\cite{PRC2013,NPA2016,FBS2014}. However, while the $\sigma$-production cross section gets a cusp, as is shown in Figs.~\ref{fig-sig} and~\ref{fig-tot}, the summed distribution remains very similar to that published in these works and still fits the data~\cite{Adl11}, provided the $\sigma$-meson parameters have the values $m_{\sigma}=297$ and $\Gamma_{\sigma}=75$ MeV. The mass of the $\sigma$ meson found in the present fit of the data~\cite{Adl13-iso} is almost the same, while the width is larger, since the ABC peak in the data~\cite{Adl13-iso} is broader. In fact, the $\sigma$-meson parameters depend on the other processes included in the calculation of the $M_{\pi\pi}$ spectrum. Its mass remains quite stable and is influenced mainly by the position of the experimental ABC peak, while the width varies stronger (but remains small compared to the free-space value of about 500 MeV). In particular, $m_{\sigma}$ varies from 290 to 320 MeV and $\Gamma_{\sigma}$ --- from about 100 to 150 MeV, when we include different combinations of the $\Delta\Delta$ and $\mathcal{D}_{12}\pi$ decay routes of the $\mathcal{D}_{03}$ resonance, the lower values corresponding to inclusion of the $\Delta\Delta$ mode only. Thus, the parameters of the $\sigma$ meson listed in Tab.~\ref{tab1} correspond to the average values.

As is seen from Fig.~\ref{fig-tot}, the calculated (summed) distributions do not reproduce the low-mass peaks in both $\pi^0\pi^0$ and $\pi^+\pi^-$ production quantitatively, but they exhibit a proper qualitative behaviour at low $M_{\pi\pi}$, which looks somewhat differently for neutral and charged dipion production. This difference cannot be reproduced without the $\sigma$-meson contribution.
The theoretical $M_{\pi\pi}$ distributions shown in Fig.~\ref{fig-tot} were calculated with the $\sigma$ total width defined by Eq.~(\ref{gst}) under the assumption that the $\sigma\pi\pi$ coupling constants obey the isospin symmetry (see Eq.~(\ref{gnc0})). By thin lines in the Figure we also show the distributions corresponding to the different coupling constants for the $\sigma\pi^0\pi^0$ and $\sigma\pi^+\pi^-$ vertices, where the difference is governed by the adjustable parameter $\alpha$ (see Eq.~(\ref{gnc})). We have achieved the quantitative description of the low-$M_{\pi\pi}$ data with $\alpha=0.23$. For this value of $\alpha$, the ratio of the coupling constants is $g^2_{\sigma c}/g^2_{\sigma n}=1.44$ instead of the usual (isospin-symmetric) value of $2$ corresponding to $\alpha=0$. In turn, the $\sigma$ partial widths are related as $\Gamma^{(0)}_{\sigma c}/\Gamma^{(0)}_{\sigma n}=1.23$ for $\alpha=0.23$ and 1.71 for $\alpha=0$. In principle, the $\sigma\pi\pi$ coupling constants should be constrained by the data on $\pi\pi$ scattering. Unfortunately, we have not found in the literature any investigation of the $\sigma\pi\pi$ coupling constants beyond the isospin symmetry even for the standard (PDG) values of the $\sigma$ mass and width. But we suppose the dynamical isospin symmetry breaking to be important for production of the near-threshold $\sigma$ meson, which undergoes the partial chiral symmetry restoration, since the pion mass difference gets more crucial for $m_{\sigma} \sim 300$ MeV than for $m_{\sigma} \sim 500$ MeV. In this case the coupling strength of the $\sigma$ meson to the neutral and charged dipions might differ substantially. At the present stage, the values obtained in this work for the $\sigma\pi\pi$ coupling constants, which govern the $\sigma\to\pi\pi$ decay widths, should be considered as a plausible phenomenology. In view of our results, a detailed microscopic investigation of this issue is highly desirable.

As an alternative source of splitting between the neutral and charged dipion production cross sections, one might consider the dynamical $\sigma$-meson generation in the final state interaction (FSI) of two pions produced via the intermediate $\mathcal{D}_{12}$ or $\Delta\Delta$ excitation. The early attempts~\cite{ABC} to describe the ABC effect by the $\pi\pi$ FSI have revealed that the isoscalar $\pi\pi$ scattering length should be 10 times larger than its experimental value $a_0 = 0.28$ fm. On the other hand, the model calculations~\cite{Roca02} have shown a substantial effect of the $\pi\pi$ FSI in the $\sigma$ channel on $\pi^0\pi^0$ photoproduction on the proton. The conclusions of the work~\cite{Roca02} are in qualitative agreement with the results obtained within the Chiral Perturbation Theory~\cite{Bernard96}, which predicted a considerable enhancement of the $\pi^0\pi^0$ photoproduction cross section near threshold due to pion loops. Both theoretical predictions~\cite{Roca02} and~\cite{Bernard96} are consistent with experimental data~\cite{Kotulla04}. Further, we have demonstrated in the present work (see Fig.~\ref{fig-sig}$b$) that the cross sections of $pn$-induced charged and neutral dipion production via the intermediate $\sigma$ meson with its free-space (PDG) parameters behave differently in the near-threshold region. Thus, while the $\sigma$ generation in the $\pi\pi$ FSI is unlikely to reproduce the total ABC enhancement, it could give some visible splitting between the $\pi^0\pi^0$ and $\pi^+\pi^-$ production cross sections, which, when added coherently to the direct $\sigma$ production mechanism from the $\mathcal{D}_{03}$ dibaryon, would be sufficient to reproduce the data on the $M_{\pi\pi}$ distributions. We postpone the detailed investigation of the $\pi\pi$ FSI effects to the future work.

\section{Energy dependence of the double-pion production cross sections}
\label{sec-ed}

\begin{figure}[!ht]
\begin{center}
\resizebox{1.0\columnwidth}{!}{\includegraphics{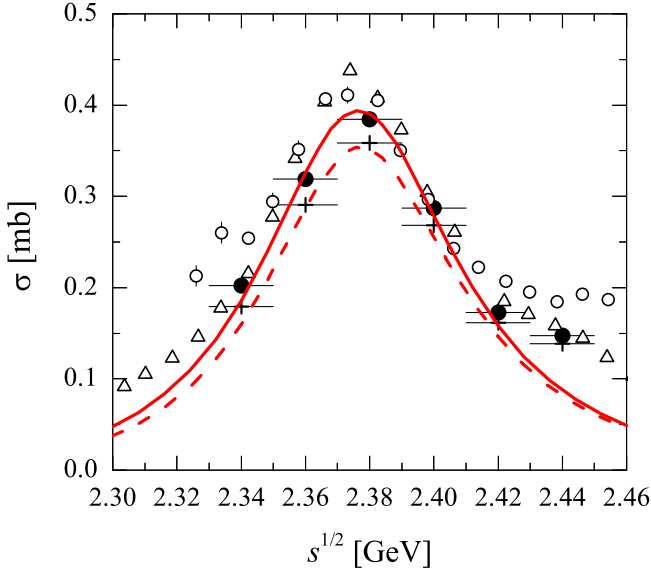}}
\end{center}
\caption{(Color online) The total cross sections in the reactions $pn \to d \pi^0\pi^0$ (multiplied by 2) and $pn \to d (\pi^+\pi^-)_0$ as functions of the invariant energy $\sqrt{s}$. The solid and dashed lines correspond to the model calculations for $\pi^0\pi^0$ and $\pi^+\pi^-$ production, respectively, including the $\mathcal{D}_{03}\to \mathcal{D}_{12}+\pi$, $\mathcal{D}_{03}\to \Delta+\Delta$, and $\mathcal{D}_{03}\to d+\sigma$ decay routes with parameters from Tab.~1 and $\alpha=0.23$ (see Eq.~(\ref{gnc})). The theoretical calculations are compared to the experimental data on $2\sigma(pn \to d \pi^0\pi^0)$ (filled circles) and $\sigma(pn \to d \pi^+\pi^-)-\textstyle{\frac{1}{2}}\sigma(pp \to d \pi^+\pi^0)$ (crosses) obtained by an integration of the respective $M_{\pi\pi}$ distributions measured in Ref.~\cite{Adl13-iso}. Also shown are the total cross section data on $2\sigma(pn \to d \pi^0\pi^0)$ from Ref.~\cite{Adl13-iso} multiplied by a factor of 0.83 (open circles) and from Ref.~\cite{Adl11} multiplied by a factor of 0.5 (open triangles) --- see Appendix.}\label{fig-st}
\end{figure}

\begin{figure}[!ht]
\begin{center}
\resizebox{0.85\columnwidth}{!}{\includegraphics{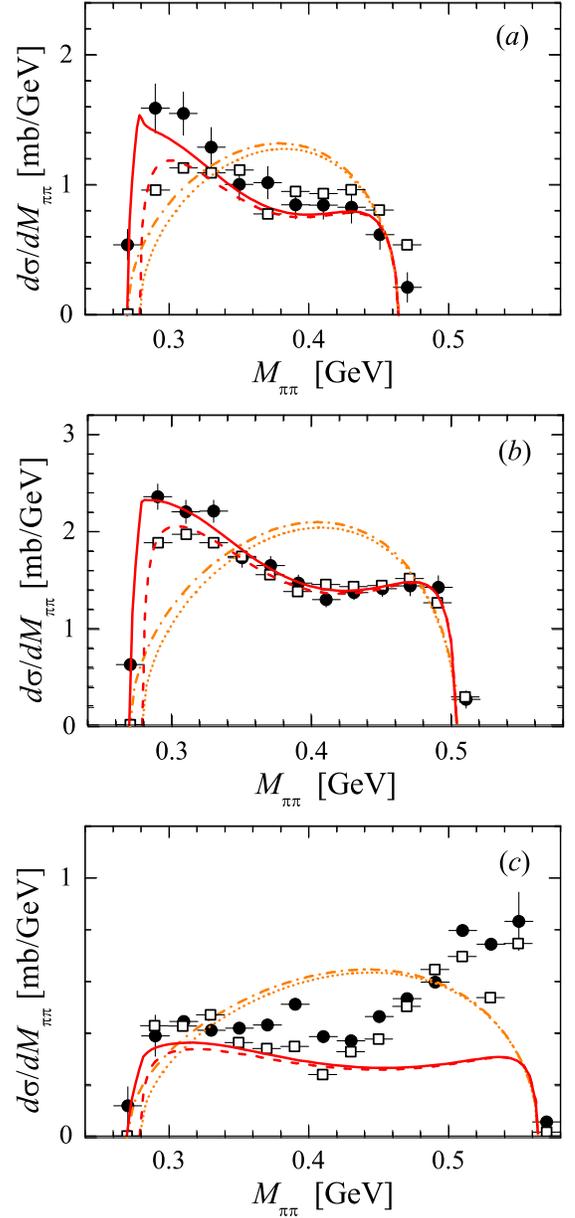}}
\end{center}
\caption{(Color online) The $\pi\pi$ invariant-mass distributions in the reactions $pn \to d \pi^0\pi^0$ (multiplied by 2) and $pn \to d (\pi^+\pi^-)_0$ at $\sqrt{s}=2.34$ GeV $(a)$, $2.38$ GeV $(b)$, and $2.44$ GeV $(c)$. The solid and dashed lines correspond to the model calculations for $\pi^0\pi^0$ and $\pi^+\pi^-$ production, respectively, including the $\mathcal{D}_{03}\to \mathcal{D}_{12}+\pi$, $\mathcal{D}_{03}\to \Delta+\Delta$, and $\mathcal{D}_{03}\to d+\sigma$ decay routes with parameters from Tab.~1 and $\alpha=0.23$ (see Eq.~(\ref{gnc})). The theoretical calculations are compared to the experimental data on $2d\sigma/dM_{\pi^0\pi^0}$ (filled circles) and $d\sigma/dM_{\pi^+\pi^-}-\frac{1}{2}d\sigma/dM_{\pi^+\pi^0}$ (open squares) taken from Ref.~\cite{Adl13-iso}. Also shown are the pure phase-space distributions for $\pi^0\pi^0$ and $\pi^+\pi^-$ production by dash-dotted and dotted lines, respectively.} 
\label{fig-ds3e}
\end{figure}

In two previous Sections we considered the $M_{\pi\pi}$ distributions in the reactions $pn \to d \pi^0\pi^0$ and $pn \to d \pi^+\pi^-$ at $\sqrt{s}=2.38$ GeV corresponding to the peak of the $\mathcal{D}_{03}$ resonance excitation. In Ref.~\cite{Adl13-iso} the $M_{\pi\pi}$ distributions and total cross sections at lower and higher energies were also measured for the above two reactions, as well as for the reaction $pp \to d \pi^+\pi^0$ in the isovector channel. In Ref.~\cite{FBS2014} we described the total cross section data~\cite{Adl13-iso,Adl11} for the $pn \to d \pi^0\pi^0$ reaction at different energies. Unfortunately, we cannot use that fit here to describe the energy dependence of the differential distributions, since the normalization of the $M_{\pi\pi}$ distributions for $\pi^0\pi^0$ and $\pi^+\pi^-$ production presented in Ref.~\cite{Adl13-iso} is not consistent with that of the total cross sections obtained in the same experiment. This inconsistency in the data~\cite{Adl13-iso} occurs mainly due to use of quasi-free scattering and different energy bins in the measurements of total and differential cross sections~\cite{Bash-PC}. These problems of the data normalization are discussed in detail in Appendix.

In Fig.~\ref{fig-st} we compare our model calculations for the total cross sections of isoscalar dipion production with the data obtained by an integration of the $M_{\pi\pi}$ distributions~\cite{Adl13-iso}.
 The rescaled total cross section data for $\pi^0\pi^0$ production are also shown. It is seen from the Figure that our model describes well the integrated $M_{\pi\pi}$ distributions for both $\pi^0\pi^0$ and isoscalar $\pi^+\pi^-$ production. It also reproduces properly the experimental trend of decreasing the isospin symmetry violation effects with the rising energy, which is related to the decrease of the low-mass enhancement.

In Fig.~\ref{fig-ds3e} our results for the $\pi\pi$ invariant-mass distributions at $\sqrt{s}=2.34$, 2.38, and 2.44 GeV are presented and compared with the data~\cite{Adl13-iso}. As in Fig.~\ref{fig-st}, we have plotted here the results obtained for $\alpha=0.23$ (see Eq.~(\ref{gnc})) to demonstrate that once the data at the resonance energy $\sqrt{s}=2.38$ GeV are reproduced by our model, the data at neighboring energies in the low-$M_{\pi\pi}$ region can also be described reasonably. Some underestimation of the data in this region is related to the contributions of other reaction mechanisms, which were not included in our model. At $\sqrt{s} = 2.34$ GeV, this is likely the $t$-channel Roper resonance $N^*(1440)$ excitation which dominates double-pion production at lower energies. Furthermore, it has been shown recently~\cite{EPJA20S,Clem20} that the $NN^*(1440)$ dibaryonic state can be formed at $\sqrt{s} \simeq 2.30$ GeV. One should bear in mind however that, due to the high $\sigma$-mesonic mode in the Roper resonance decay~\cite{PDG20}, a similar mechanism for the isospin symmetry breaking to that we propose here for the $\mathcal{D}_{03}$ decay may be applicable also in the region of the Roper resonance (or $NN^*(1440)$ dibaryon) dominance. On the other hand, at higher energies close to the $\Delta\Delta$ threshold, the high-$M_{\pi\pi}$ peak arises, which can likely be reproduced by the $t$-channel $\Delta\Delta$ process not included in our current framework. Nevertheless, the $\mathcal{D}_{03}$ contribution in our model dominates the low-$M_{\pi\pi}$ region at these energies as well. Again, we see from Fig.~\ref{fig-ds3e} that our calculations properly reflect the decrease of the near-threshold isospin symmetry breaking effects with the decrease of the ABC peak at higher energies which is clearly seen in the data on $M_{\pi\pi}$ distributions. Adding the $t$-channel $\Delta\Delta$ process should lead to further restoration of the isospin symmetry in the differential and total cross sections.
 Thus, the observed isospin symmetry breaking in the region of the $\mathcal{D}_{03}$ excitation appears to be intimately related to the ABC peak. Both these effects are explained in our model as a consequence of the intermediate near-threshold $\sigma$-meson production.

 It is known however that the ABC peak is very moderate, if present at all, in the double-pion production reactions with the unbound $pn$ pair in the final state~\cite{Adl15-pn2pi}. Within our model, this can be explained as follows. Since we consider here the $\sigma$-meson emission via the $\mathcal{D}_{03} \to \mathcal{D}_{01}+\sigma$ decay, which is a transition between two six-quark states, the contribution of this mechanism to the reaction with the $d\pi\pi$ (or $pn\pi\pi$) final state is related to the weight of the six-quark $\mathcal{D}_{01}$ component in the final deuteron (or $pn$ pair). Our preliminary calculations show that the weight of the compact six-quark state in the bound deuteron is much larger than that in the $pn$ continuum~\cite{PR-pn2pi}. So, the $\sigma$-meson emission will be dynamically suppressed in case of the $pn \to pn(\pi\pi)_0$ reactions. This suppression will not lead to a significant reduction of the total cross section (which should be about 15\% smaller in case of the unbound $pn$ pair in the final state~\cite{Bash15-BR}), since the $\sigma d$ branch in our model is less than 5\% of the $pn\to d(\pi\pi)_0$ cross section, while the summed contribution of other mechanisms is about 80\%. The detailed calculations of the $pn \to pn(\pi\pi)_0$ reactions are in progress.

\section{Summary and outlook}
\label{sec-sum}

We have shown that the observed suppression of the near-threshold enhancement (the so-called ABC effect) in the $\pi\pi$ invariant-mass spectrum in the reaction $pn \to d\pi^+\pi^-$ compared to that in the reaction $pn \to d\pi^0\pi^0$ can be at least partially explained by the intermediate $\mathcal{D}_{03}(2380)$ (denoted also as $d^*(2380)$) dibaryon decay with the scalar $\sigma$-meson emission. The same mechanism is capable to explain the appearance of the ABC effect itself~\cite{PRC2013}, provided the $\sigma$ mass and width are shifted downwards to the values of about $m_{\sigma}=290$--$320$ MeV and $\Gamma_{\sigma}=75$--$150$ MeV due the partial chiral symmetry restoration in the excited $\mathcal{D}_{03}$ dibaryon. Being a near-threshold resonance, such a renormalized $\sigma$ meson produces a cusp in the $\pi^0\pi^0$ production cross section at the $\pi^+\pi^-$ threshold, thus giving the visible splitting between the neutral and charged dipion production cross sections in the near-threshold region. The free-space $\sigma$ meson with the parameters $m_{\sigma}\simeq\Gamma_{\sigma}\simeq 500$ MeV produces a similar (though less prominent) cusp but a different shape of the $M_{\pi\pi}$ distribution, which peaks at the nominal $\sigma$ mass.

Other mechanisms proposed for double-pion production in $pn$ collisions in the region of the $\mathcal{D}_{03}(2380)$ excitation, such as its decay into the $\pi+\mathcal{D}_{12}(2150)$ or $\Delta+\Delta$ intermediate states, exhibit no isospin symmetry breaking effects except for a shift of the low-mass peak in the $\pi^+\pi^-$ production channel due to the phase space reduction for the charged dipions. It is not surprising, since both $\mathcal{D}_{12}$ and $\Delta$ resonances are located far from the respective pion production thresholds, so the dynamics of their decay is almost unsensitive to the 5-MeV mass difference between the neutral and charged pions.

However, when the intermediate $\sigma$ emission is added coherently to the $\mathcal{D}_{12}$ or $\Delta\Delta$ production, the contribution of the former mechanism gets reduced. Then the additional sources for the isospin symmetry breaking should be considered, such as dynamical $\sigma$ generation in the $\pi\pi$ FSI or the different coupling strength of the $\sigma$ meson to the neutral and charged dipions. These additional mechanisms are also related to the $\sigma$-meson production.

 The intermediate $\sigma$-meson excitation mechanism might explain the similar isospin symmetry breaking effects in the reactions $pd \to {}^3{\rm He}\pi\pi$ and $pp \to pp\pi\pi$ in the GeV energy region as shown up in experiments of the CELSIUS/WASA Collaboration~\cite{Bash06}. In $pp$ collisions, the $\sigma$ meson can be emitted from the intermediate isovector dibaryons as was claimed in~\cite{NPA2016}. Further, the recent work~\cite{EPJA20S} has demonstrated a crucial role of the dibaryons (both isovector and isoscalar) located near the $NN^*(1440)$ threshold in elastic and inelastic $S$-wave $NN$ scattering. In particular, a clear indication of such a dibaryon formation has been found in the data on $pp$-induced two-pion production~\cite{Skorodko09}. This isovector dibaryon should decay predominantly via the $NN^*(1440)$ intermediate state, and the Roper resonance $N^*(1440)$ is known to have a very strong $\sigma N$ decay mode~\cite{PDG20}.

 In this regard, it is also worth mentioning another CELSIUS/WASA experiment~\cite{Bash05} on the reaction $pp \to pp \gamma\gamma$, which clearly showed a cusp in the $\gamma\gamma$ spectrum at the two-pion threshold. This cusp was interpreted~\cite{Bash05} as being due to opening of the $\pi\pi$ channel in the decay of an intermediate $\sigma$ meson with a mass $M_{\sigma} \simeq 300$
MeV. An indication of the very light $\sigma$-meson generation in both $\pi\pi$ and $\gamma\gamma$ production in $dC$ collisions has been also found in the experiments of the Dubna group~\cite{Abraamyan}.


It should be stressed that dynamics of light scalar meson production in hadronic collisions is poorly understood to date. Theoretical predictions and experimental indications of $\sigma$-meson production in $NN$ collisions as well as in quarkonia decays at high energies can be found, e.g., in Refs.~\cite{Kisslinger05,Alde97,Ablikim07}. The results of the present work suggest that both ABC effect and near-threshold isospin symmetry violation in the $\pi\pi$ invariant-mass spectra in $NN$-induced double-pion production indicate the $\sigma$-meson generation in $NN$ collisions at intermediate energies as well. At last, we should emphasise that any reliable confirmation for the near-threshold $\sigma$-meson production with the reduced mass and width (with respect to their free-space values) should be crucially important for the validity of the novel dibaryon concept for the short-range nuclear force, where the generation of such an intermediate $\sigma$-meson with the low mass $m_{\sigma}=300$--$350$ MeV plays a key role~\cite{JPG01K,IJMP02K,AP10K}.

\textit{To summarize,} the observed isospin symmetry breaking in double-pion production in $NN$, $Nd$, etc., collisions, which is manifested in the suppression of the $\pi^+\pi^-$ production cross section in comparison to the $\pi^0\pi^0$ one near the two-pion threshold, gives a strong argument in favor of the generation of the intermediate light scalar $\sigma$ mesons in such processes. The $\sigma$ mesons are likely to be emitted directly from the intermediate dibaryon resonances. This brings support to the $\sigma$-dressed dibaryon mechanism for the short-range $NN$ interaction as proposed in Refs.~\cite{JPG01K,IJMP02K,AP10K}. The recent experimental and theoretical confirmations of this novel mechanism can be found in Refs.~\cite{EPJA20S,PLB20D,PRD20P,YAF19}.


\section*{ACKNOWLEDGEMENTS} We are indebted to Dr. M. Bashkanov and Prof. H. Clement for the valuable comments on the WASA-at-COSY experimental results. We also appreciate the fruitful discussion with Prof. C. Wilkin on the isospin symmetry violation in nucleon-nucleon collisions.
The work was done under partial financial support from the Russian Foundation for Basic Research, grants Nos.~19-02-00011 and 19-02-00014, and the Foundation for the Advancement of Theoretical Physics and Mathematics ``BASIS''.

\appendix
\section*{Appendix: Normalization of experimental data}
\label{app}

From the analysis of experimental data~\cite{Adl13-iso}, we found that the normalization of the $M_{\pi\pi}$ distributions presented in this work is not consistent with that of the total cross sections measured in the same experiment. In fact, we found that the total cross sections obtained by an integration of the $M_{\pi\pi}$ distributions presented in Ref.~\cite{Adl13-iso} at $\sqrt{s}=2.34$--$2.44$ GeV are lower than the respective total cross section data for $\pi^0\pi^0$ production at all measured energies (by a factor of 1.2--1.35) and for $\pi^+\pi^-$ production at energies $\sqrt{s} < 2.38$ GeV (by a factor of 1.3--1.4). At the same time, the normalization of the differential and total cross section data is consistent for $\pi^+\pi^-$ production at $\sqrt{s} \geq 2.38$ GeV and for $\pi^+\pi^0$ production at all energies.

\begin{figure*}[!ht]
\begin{center}
\resizebox{1.0\textwidth}{!}{\includegraphics{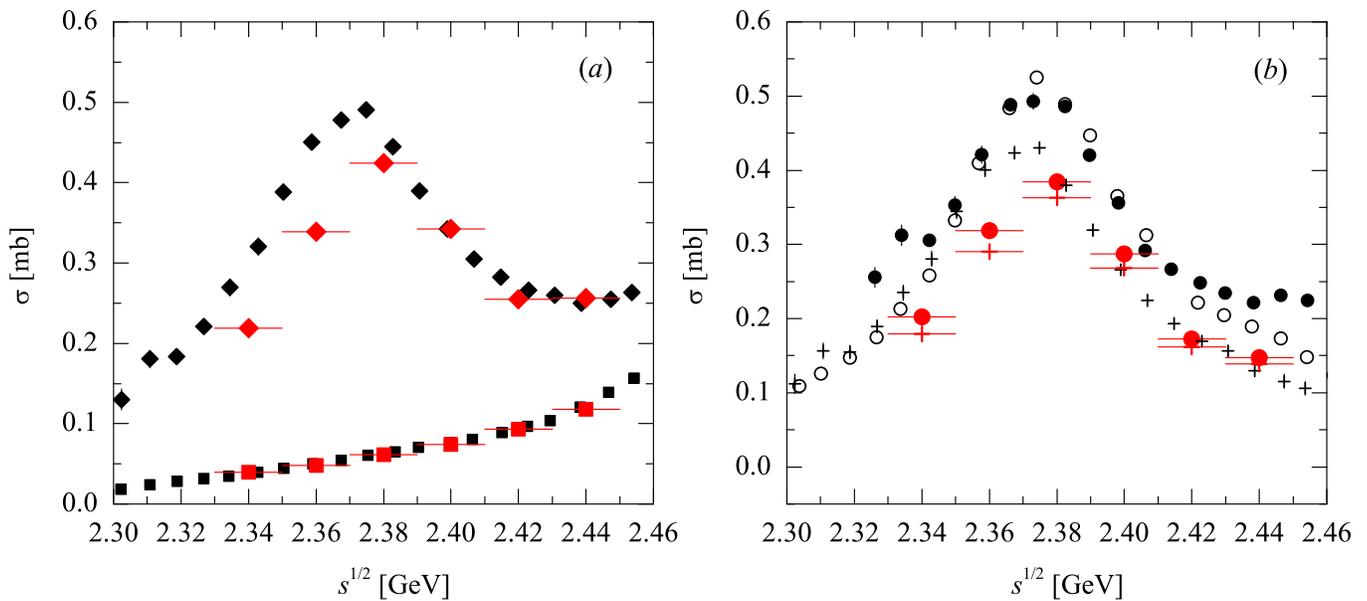}}
\end{center}
\caption{(Color online) $(a)$ The total cross section data $\sigma(pn \to d \pi^+\pi^-)$ (diamonds) and $\textstyle{\frac{1}{2}}\sigma(pp \to d \pi^+\pi^0)$ (squares) from Ref.~\cite{Adl13-iso}. The same symbols with horizontal error bars show the total cross sections obtained by an integration of the respective $M_{\pi\pi}$ distributions measured in~\cite{Adl13-iso}. $(b)$ The total cross section data $2\sigma(pn \to d \pi^0\pi^0)$ from Refs.~\cite{Adl13-iso} (filled circles) and~\cite{Adl11} (open circles, multiplied by 0.6) and $\sigma(pn \to d \pi^+\pi^-)-\textstyle{\frac{1}{2}}\sigma(pp \to d \pi^+\pi^0)$ from Ref.~\cite{Adl13-iso} (crosses). The same symbols with horizontal error bars show the total cross sections obtained by an integration of the respective $M_{\pi\pi}$ distributions measured in~\cite{Adl13-iso}.} \label{fig-stnorm}
\end{figure*}

In Fig.~\ref{fig-stnorm} we illustrate the above normalization problems for the data~\cite{Adl13-iso}. In the Figure, the total cross sections for three reactions $pn \to d \pi^0\pi^0$, $pn \to d \pi^+\pi^-$, and $pp \to d \pi^+\pi^0$ measured in Ref.~\cite{Adl13-iso}, as well as the total cross section for isoscalar $\pi^+\pi^-$ production, i.e., $\sigma(\pi^+\pi^-)_0=\sigma(\pi^+\pi^-)-\textstyle{\frac{1}{2}}\sigma(\pi^+\pi^0)$, are shown. The total cross section data~\cite{Adl11} for $\pi^0\pi^0$ production are also shown in Fig.~\ref{fig-stnorm}$b$ (these data have been multiplied by a factor of 0.6 for consistency with the data~\cite{Adl13-iso}). The experimental total cross sections are compared with the values obtained by an integration of the respective $M_{\pi\pi}$ distributions presented in Ref.~\cite{Adl13-iso}.

The above inconsistency in the data occurs mainly due to the use of different energy bins in the measurements of total and differential cross sections~\cite{Bash-PC}. When using quasi-free $pn$ scattering, the obtained distributions should be corrected for the rapid flux variation within the energy bins, and it was not done in Ref.~\cite{Adl13-iso}. So, the absolute normalization of the total cross sections is more reliable than that of the $M_{\pi\pi}$ spectra, since narrower energy bins were used for the total cross section measurements. The large systematic errors in the measured $M_{\pi\pi}$ distributions in the near-threshold region as well as averaging the distributions over 10-MeV bins in $M_{\pi\pi}$ also complicate the correct data normalization~\cite{Bash-PC}. For the same reasons, the low-mass peak in the data~\cite{Adl13-iso} occurred to be lower and broader than that obtained in Ref.~\cite{Adl11} for $\pi^0\pi^0$ production. The overall quality of the data~\cite{Adl13-iso} is therefore not as good as that of the older data~\cite{Adl11}. On the other hand, only Ref.~\cite{Adl13-iso} provides the data for all three double-pion production channels at the same energies and allows for an analysis of the isospin symmetry breaking in these reactions. The differential distributions measured in this work seem to be properly related to each other (at the given energy), since they exhibit the strong isospin symmetry breaking in the near-threshold region which vanishes at high $M_{\pi\pi}$. The total cross sections obtained by an integration of the differential $M_{\pi\pi}$ distributions also appear to have a correct energy dependence (at least relatively to each other), since the isospin symmetry violation should get weaker at higher energies (cf. red circles and crosses with horizontal error bars in Fig.~\ref{fig-stnorm}$b$). This is substantiated by the fact that the high-mass region prevails in the isoscalar dipion production cross sections when the energy rises up to the $\Delta\Delta$ threshold. On the other hand, the total cross section data~\cite{Adl13-iso} exhibit just the opposite trend (cf. black circles and crosses in Fig.~\ref{fig-stnorm}$b$). This is likely related to the 10--20\% overall uncertainty in the total cross sections normalization~\cite{Bash-PC}. Therefore, in the present work we take the $M_{\pi\pi}$ distributions from Ref.~\cite{Adl13-iso} as they are and compare our model calculations for the total production cross sections with the integrated $M_{\pi\pi}$ distributions rather than the total cross section data~\cite{Adl13-iso}.

In view of the above problems, it is nontrivial to rescale the $M_{\pi\pi}$ distributions measured in Ref.~\cite{Adl13-iso} to make their normalization consistent with that of the total cross sections. In fact, the $M_{\pi\pi}$ spectra can be scaled by some factor, but this factor should be {\em the same for all three reactions} to keep the relation between the cross sections for the different isospin channels. By minimising the $\chi^2$ for the data for all three reactions at all measured energies (including also the older data~\cite{Adl11}), we found that the differential $M_{\pi\pi}$ distributions presented in Ref.~\cite{Adl13-iso} should be multiplied by a factor of 1.2 to get the average consistency with the total cross section data~\cite{Adl13-iso}.
On the other hand, the total cross section data can be also renormalized by a factor of $10$--$20$\% corresponding to an overall uncertainty in their absolute normalization~\cite{Bash-PC}. In Fig.~\ref{fig-st} we plotted the total cross section data~\cite{Adl13-iso} for isoscalar dipion production multiplied by $1/1.2=0.83$, thus making them much closer to the integrated $M_{\pi\pi}$ distributions than the initial data~\cite{Adl13-iso}. As was also shown in Ref.~\cite{Adl13-iso}, the older data~\cite{Adl11} for both differential and total $\pi^0\pi^0$ production cross sections should be renormalized by a factor of about 0.6 for consistency with the data~\cite{Adl13-iso} on the total cross sections. In Figs.~\ref{fig-d12}, \ref{fig-dd}, and \ref{fig-st} we have additionally decreased the data~\cite{Adl11} for consistency with the normalization of the differential $M_{\pi\pi}$ distributions measured in~\cite{Adl13-iso}. Thus, the average renormalization factor for the data~\cite{Adl11} is $0.6/1.2=0.5$ (see Fig.~\ref{fig-st}), while the particular renormalization factor at $\sqrt{s}=2.38$ GeV is $0.6/1.32=0.45$ (see Figs.~\ref{fig-d12} and \ref{fig-dd}).

\end{document}